\documentclass[%
 reprint,
 superscriptaddress,
 amsmath,amssymb,
 aps,
prb,
]{revtex4-2}

\usepackage[colorlinks,bookmarks=false,citecolor=darkblue,linkcolor=red,urlcolor=blue]{hyperref}

\usepackage{graphicx}
\usepackage{dcolumn}
\usepackage{bm}
\usepackage{multirow}
\usepackage{booktabs}

\usepackage[dvipsnames]{xcolor}

\usepackage{boldline}

\usepackage{color}
\usepackage[colorlinks,bookmarks=false,citecolor=darkblue,linkcolor=red,urlcolor=blue]{hyperref}
\definecolor{darkblue}{rgb}{0,0.02,0.45}

\newcommand{\fmo}{Fe$_{2}$Mo$_3$O$_8$}

\newcommand{\TN}{\ensuremath{T_{\mathrm{N}}}}

\usepackage[warn]{textcomp}
\usepackage{mathtools}
\DeclarePairedDelimiter\bra{\langle}{\rvert}
\DeclarePairedDelimiter\ket{\lvert}{\rangle}
\DeclarePairedDelimiterX\braket[2]{\langle}{\rangle}{#1 \delimsize\vert #2}

\usepackage[version=4]{mhchem}
\usepackage{mathptmx}

\begin{document}

\title{Optical magnetoelectric effect in the polar honeycomb antiferromagnet Fe$_2$Mo$_3$O$_8$}
\date{\today}

\author{K.~V.~Vasin}
\affiliation{Experimental Physics V, Center for Electronic
Correlations and Magnetism, Institute for Physics, University of Augsburg, D-86135 Augsburg, Germany}
\affiliation{Institute for Physics, Kazan (Volga region) Federal University, 420008 Kazan, Russia}
\author{A.~Strini\'{c}}
\affiliation{Experimental Physics V, Center for Electronic
Correlations and Magnetism, Institute for Physics, University of Augsburg, D-86135 Augsburg, Germany}
\author{F.~Schilberth}
\affiliation{Experimental Physics V, Center for Electronic
Correlations and Magnetism, Institute for Physics, University of Augsburg, D-86135 Augsburg, Germany}
\author{S.~Reschke}
\affiliation{Experimental Physics V, Center for Electronic
Correlations and Magnetism, Institute for Physics, University of Augsburg, D-86135 Augsburg, Germany}
\affiliation{TOPTICA Photonics AG, Lochhamer Schlag 19, 82166 Gr\"afelfing, Germany}
\author{L.~Prodan}
\affiliation{Experimental Physics V, Center for Electronic
Correlations and Magnetism, Institute for Physics, University of Augsburg, D-86135 Augsburg, Germany}
\author{V.~Tsurkan}
\affiliation{Experimental Physics V, Center for Electronic
Correlations and Magnetism, Institute for Physics, University of Augsburg, D-86135 Augsburg, Germany} 
\affiliation{Institute of Applied Physics, Moldova State University, MD-2028~Chi\c{s}in\u{a}u, Moldova}
\author{A.~R.~Nurmukhametov}
\affiliation{Institute for Physics, Kazan (Volga region) Federal University, 420008 Kazan, Russia}
\author{M.~V.~Eremin}
\affiliation{Institute for Physics, Kazan (Volga region) Federal University, 420008 Kazan, Russia}
\author{I.~K\'ezsm\'arki}
\affiliation{Experimental Physics V, Center for Electronic
Correlations and Magnetism, Institute for Physics, University of Augsburg, D-86135 Augsburg, Germany}
\author{J.~Deisenhofer}
\affiliation{Experimental Physics V, Center for Electronic
Correlations and Magnetism, Institute for Physics, University of Augsburg, D-86135 Augsburg, Germany}

\date{\today}

\begin{abstract}
The lack of both time-reversal and spatial inversion symmetry in polar magnets is a prerequisite for the occurrence of optical magnetoelectric effects such as nonreciprocal directional dichroism with the potential for the realization of optical diodes. In particular,  antiferromagnetic materials with magnetic excitations in the THz range such as \fmo{} are promising candidates for next-generation spintronic applications. In a combined experimental and theoretical effort we investigated the THz excitations of the polar honeycomb antiferromagnet \fmo{} in external magnetic fields and their  nonreciprocal directional dichroism, together with the temperature dependence of the electronic transitions in the mid- and near-infrared frequency range. Using an advanced single-ion approach for the Fe ions, we are able to describe optical excitations from the THz to the near-infrared frequency range quantitatively and successfully model the observed nonreciprocal directional dichroism in the THz regime.
\end{abstract}

\maketitle

\section{Introduction}
The versatility of multiferroic compounds with cross correlations of spin and charge degrees of freedom \cite{Spaldin:2019,Dong:2019} makes them a rich playground to explore new materials for applications.  Antiferromagnetic materials came into focus for spintronic application, as their elementary magnetic excitations are often found in the THz frequency range \cite{Jungwirth:2018,Baltz:2018,Nemec:2018}. The lack of both spatial and time-inversion symmetry in many of these materials allows the observation of novel optical magnetoelectric phenomena, 
of which the nonreciprocal directional dichroism of light is among the most exciting \cite{Rikken:2002,Jung:2004,Saito:2008,Kezsmarki:2011,Bordacs:2012,Takahashi:2013,Szaller:2013,Kezsmarki:2014,Szaller:2014,Kuzmenko:2015,Kezsmarki:2015,Bordacs:2015,Toyoda:2015,Kurumaji:2017b,Iguchi:2017,Kocsis:2018,Yu:2018,Szaller:2019,Viirok:2019,Kimura:2020,Yokosuk:2020,Vit:2021,Nikitchenko:2021,Kopteva:2022,Reschke:2022}, as the difference in absorption for the two counter-propagating light beams can, in the extreme limit, lead to one-way transparency \cite{Kezsmarki:2014,Kurumaji:2017} in these materials and facilitate the construction of optical-diode devices \cite{Yu:2018}. Albeit, theoretical approaches to quantitatively describe the observed nonreciprocal directional dichroism effects are rare, focusing on density-functional-theory approaches for the multiferroic Ni$_3$TeO$_6$ at energies above 1~eV \cite{Yokosuk:2020,Park:2022}. 

\begin{figure}[htb]
    \centering
    \includegraphics[width=1\linewidth,clip]{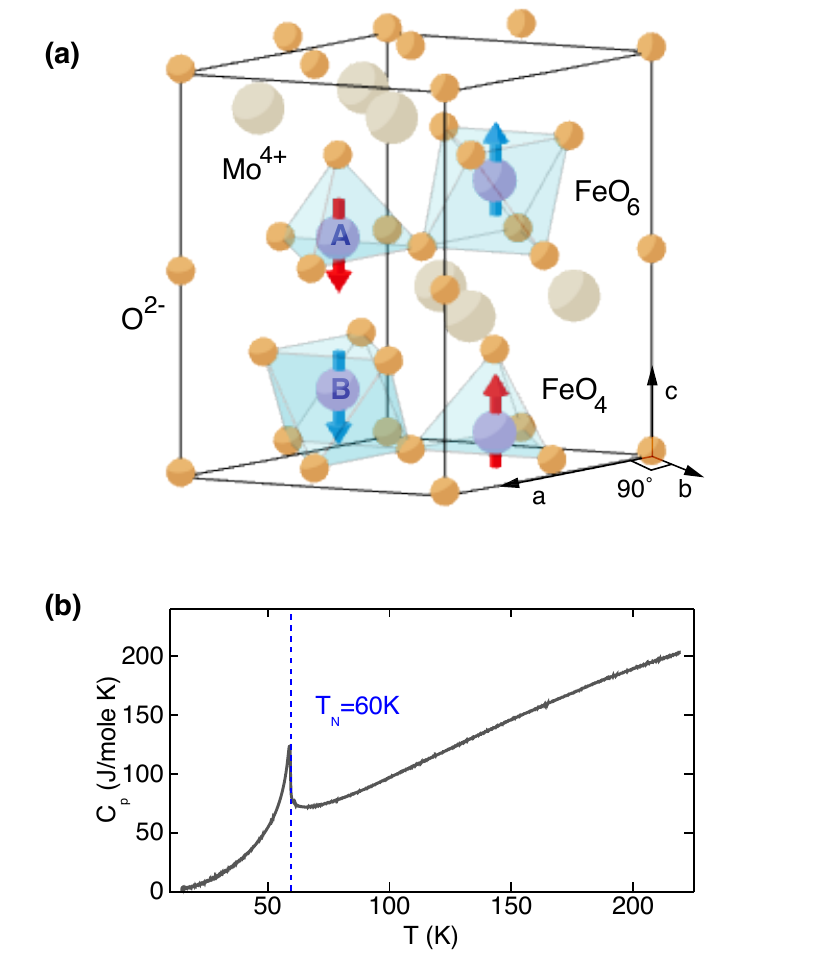}
    \caption{\label{fig:structure}  (a) Unit cell of \fmo{} in  the antiferromagnetic state with tetrahedral (A) and octahedral (B) sites shown in red and blue, respectively, and the polarization along the $c$-axis. The  depicted \textit{b}-axis is perpendicular to the hexagonal $a$-axis. (b) Temperature dependence of the specific heat $C_p$ showing the antiferromagnetic ordering at $\TN{}=60$\,K.}
\end{figure}

The material \fmo{} belongs to the family of polar molybdenum oxides $A_{2}$Mo$_3$O$_8$  ($A$ = Mn, Fe, Co, Ni, Zn), which has recently come into the focus of research, as different magnetically ordered ground states can be formed and tuned by external magnetic fields or doping \cite{Kurumaji:2015,Wang:2015,Kurumaji:2017,Tang:2019,Csizi:2020,Tang:2021,Tang:2022,Prodan:2022}. The $A^{2+}$ ions occupy corner-sharing tetrahedral (A) and octahedral (B) sites and are responsible for the magnetism, while the Mo ions build non-magnetic trimers \cite{Varret:1972,Cotton:1964,Wang:2015}. \fmo{}  has a hexagonal structure with the polar space group $P6_3mc$, the polarization along the $c$-axis, and a collinear antiferromagnetic order below $\TN{}=60$\,K \cite{Bertrand:1975,McAlister:1983} (see Fig.~\ref{fig:structure}). Recently, it was  identified to fulfill the criteria to be called an altermagnet, too \cite{Cheong2024}.  Upon applying a magnetic field and/or substitution of Fe by nonmagnetic Zn, a transition to a ferrimagnetic state can be induced in \fmo{} \cite{Wang:2015,Kurumaji:2015}, which exhibits very unusual magnetization reversal properties \cite{Ghara:2023}. Furthermore, a  giant thermal Hall effect has been reported and ascribed to magnon-phonon coupling effects \cite{Ideue:2017,Park:2020}. The optical excitation spectra of pure and Zn-doped \fmo{} have been investigated beforehand, including electromagnons in the THz range \cite{Kurumaji:2017a,Kurumaji:2017b,Yu:2018,Csizi:2020}, lattice and electronic excitations in the infrared frequency range \cite{Stanislavchuk:2020,Reschke:2020,Park:2021}, and optically induced magneto-optical Kerr effects \cite{Sheu:2019}. Recently, Raman and neutron scattering studies revealed the presence of magnon-phonon coupling effects and a hybrid nature of the low-lying excitations in \fmo{} \cite{Wu:2023,Bao:2023}.

Theoretical efforts to explain the excitation spectra of \fmo{} have either focused on  low-energy spin Hamiltonians  \cite{Wu:2023,Bao:2023} or on single-ion crystal-field calculations \cite{Stanislavchuk:2020} of the Fe energy levels, but a real comprehensive model of the optical excitations including matrix elements and selection rules remained evasive and set the starting point for our quest to  understand and model the optical excitations and the nonreciprocal directional dichroism in \fmo. Therefore, we investigated the optical excitations in the THz regime in an external magnetic field  and in the mid-infrared frequeny range in zero-field. In addition, we studied directional dichroism effects for the THz excitations in the antiferromagnetically ordered phase. To model all our observations quantitatively, we used an advanced theoretical single-ion approach for the Fe$^{2+}$ ions in \fmo{}  and took into account the effects of an external magnetic-field. Thus, we assigned the origin of the mid- and near-infrared excitations to the corresponding Fe sites and successfully modelled the observed nonreciprocal directional dichroism effects of the low-lying THz modes by the same approach with a unified set of material parameters.

\begin{figure}[htb]
    \centering
    \includegraphics[width=0.8\linewidth,clip]{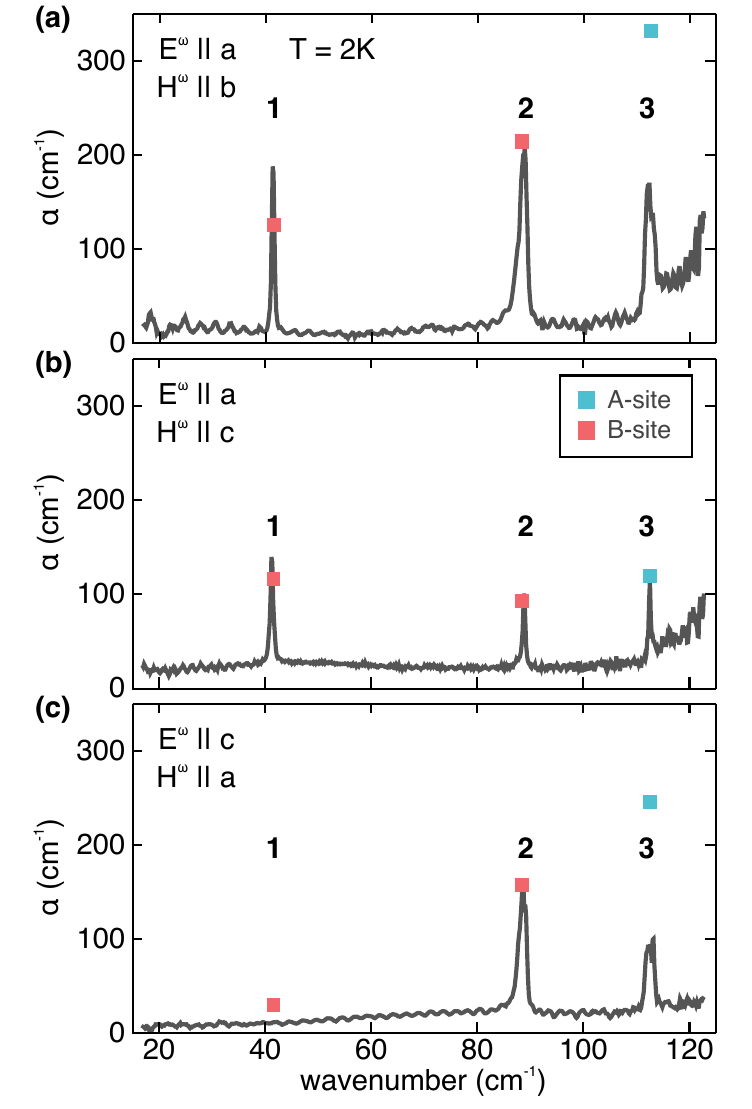}
    \caption{\label{fig:specs}  Absorption coefficient in the THz range in \fmo{} for the three orthogonal light polarization configurations (a) $E^\omega \parallel  a$ \& $H^\omega \parallel  b$, (b) $E^\omega \parallel  a$ \& $H^\omega \parallel c$, and (c) $E^\omega \parallel c$ \& $H^\omega \parallel  a$ at 2\,K. The solid squares indicate the calculated absorption coefficient as described in the text, red refers to an A-site related excitation, blue to a B-Site related one.}
\end{figure}

\section{Experimental Details}
Single crystals of \fmo{} were grown by the
chemical transport reaction method at temperatures between 950 and
900$\,^{\circ}$C. TeCl$_4$ was used as the source of the transport
agent (see \cite{Reschke:2020} for details). X-ray diffraction revealed a single-phase composition with a hexagonal symmetry using space group $P6_3mc$ and lattice constants 
$a=5.773(2)$\,\AA{} and $c=10.017(2)$\,\AA{}. The samples were characterized by specific heat (shown in Fig.~\ref{fig:structure}(b)) and magnetization (not shown) measurements, which clearly show the antiferromagnetic transition at $T_N=60$~K in agreement with literature \cite{Kurumaji:2015,Wang:2015}.

Temperature and magnetic field dependent time-domain THz spectroscopy measurements were performed on  plane-parallel $ab$- and $ac$-cut single crystals of \fmo{}. A Toptica TeraFlash time-domain THz spectrometer was used in combination with a superconducting magnet, which allows for measurements at temperatures down to 2\,K and in magnetic fields up to $\pm 7$\,T. Transmission measurements were performed in Voigt configuration with the magnetic field parallel or perpendicular to the $c$-axis. In the infrared frequency regime, reflectivity measurements were performed using a Bruker
Fourier-transform IR-spectrometer IFS 66v/S and transmission measurements  were performed using a Bruker Vertex 80v equipped with a IR Microscope Hyperion 2000.

\section{Experimental Results}

\subsection{\label{sec:thzmodes}THz modes - selection rules and magnetic-field dependence}

\begin{figure*}[htb]
    \centering
    \includegraphics[width=\linewidth,clip]{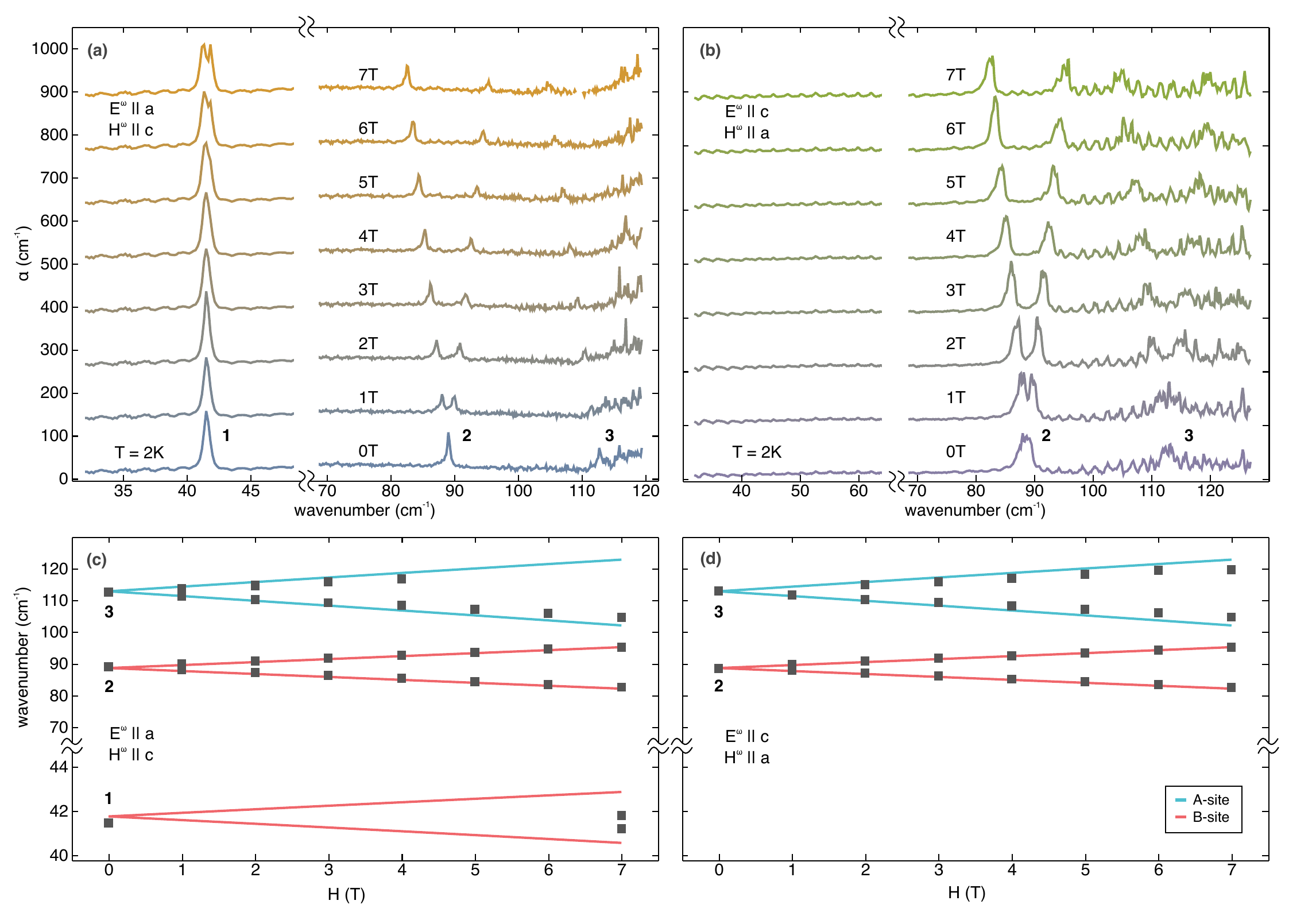}
    \caption{\label{fig:fdwide} Magnetic field dependence of the THz absorption spectra at 2~K with the external field $\mathbf{H}\parallel c$ for polarization configurations (a) $E^\omega \parallel  a$ \& $H^\omega \parallel c$ and (b) $E^\omega \parallel c$ \& $H^\omega \parallel  a$. Spectra in panels (a) and (b) are shifted with respect to the zero-field spectrum for clarity. Panels (c) and (d) show a comparison of the experimental eigenfrequencies (solid squares) for the spectra in (a) and (b), with the calculated field dependence (solid lines) as described in the text. Red lines refer to  A-site related excitations, blue lines to a B-Site related one. }
\end{figure*}

 In Figs.~\ref{fig:specs}(a)-(c) we show the THz absorption spectra for all three polarisation configurations at 2~K. The three absorption peaks at 41\,cm$^{-1}$ (mode \textbf{1}), 89\,cm$^{-1}$ (mode \textbf{2}), and 112\,cm$^{-1}$ (mode \textbf{3}) appear only below  $T_\mathrm{N}$ and identify the characteristic low-energy excitations of the antiferromagnetically ordered state. While the former two modes were observed in a previous study \cite{Kurumaji:2017a}, mode \textbf{3} at 112\,cm$^{-1}$ has not been reported before in pure \fmo{}.  In lightly Zn doped \fmo{}, however, a corresponding excitation at 114\,cm$^{-1}$ has been reported \cite{Csizi:2020}. From the three different polarization orientations, the electric-dipole selection rule $E^{\omega}\parallel a$ \cite{Kurumaji:2017a} could be confirmed for mode \textbf{1}. The other two modes are visible in all three polarization configurations, i.e. they are both electric and magnetic-dipole active and, hence, are candidates for optical magneto-electric effects such as the non-reciprocal directional dichroism (see below). 
 
 \begin{table}[b]
\squeezetable
\begin{ruledtabular}
\centering\footnotesize
   \begin{tabular}{cccccc}
\multirow{3}{*}{\textbf{mode}} & \multicolumn{2}{c}{$\boldsymbol{\omega_0}$} & \textbf{$\boldsymbol{ab}$-cut}                                                                          & \multicolumn{2}{c}{\textbf{$\boldsymbol{ac}$-cut}}                                         \\
& \multicolumn{2}{c}{$[\mathrm{cm}^{-1}]$}                                                   & \multirow{2}{*}{\begin{tabular}[c]{@{}c@{}}$E^\omega\parallel a $ \\$H^\omega\parallel b$\end{tabular}} & \multicolumn{1}{c}{\multirow{2}{*}{\begin{tabular}[c]{@{}c@{}}$E^\omega\parallel a $ \\$H^\omega\parallel c$\end{tabular}}} & \multirow{2}{*}{\begin{tabular}[c]{@{}c@{}}$E^\omega\parallel c $ \\$H^\omega\parallel a$\end{tabular}}  \\
                                         
     & exp   & calc      & & \multicolumn{1}{c|}{}       &       \\ \midrule
\textbf{\textbf{1}}            & 41.5  & 41.8                                                            & $\checkmark$                                                                                            & $\checkmark$                                                                                                                 & $\times$                                                                                                 \\
\textbf{2}                     & 89.0  & 88.6                                                            & $\checkmark$                                                                                            & $\checkmark$                                                                                                                 & $\checkmark$                                                                                             \\
\textbf{3}                     & 112.6 & 112.8                                                           & $\checkmark$                                                                                            & $\checkmark$                                                                                                                 & $\checkmark$                                                                                             \\ \midrule
$EM$                           & 40    &  -                                                               & $\checkmark$                                                                                            & $\checkmark$                                                                                                                 & $\times$                                                                                                 \\
$MM_1$                         & 90    &     -                                                            & $\checkmark$                                                                                            & $\times$                                                                                                                     & $\checkmark$                                                                                            
\end{tabular}
    \caption{\label{tab:selectionrules} Upper part: Selection rules for the
    excitations found in \fmo{} at 2~K for the different polarization configurations
    as shown in Fig. \ref{fig:specs}. The  notation $\checkmark$
    and $\times$ indicates the presence or absence of a mode. Lower
    part: reported excitation frequencies and selection rules in the AFM phase of
    \fmo{} at 4.5~K taken from Ref.~\cite{Kurumaji:2017a}.}
\end{ruledtabular}
\end{table}

 While mode \textbf{1} exhibits a similar linewidth for both active polarization configurations, the absorption peaks for modes \textbf{2} and \textbf{3} are narrower for $E^\omega \parallel  a$ \& $H^\omega \parallel c$, which suggests that additional relaxation channels are active for $H^\omega \parallel a,b$. Note that mode \textbf{2} had not been observed previously for $E^\omega \parallel a$ \& $H^\omega \parallel c$ \cite{Kurumaji:2017a}, which lead to an assignment as being only magnetic-dipole active. The modes and their selection rules are summarized in Table~\ref{tab:selectionrules} and compared to the results of Ref.~\cite{Kurumaji:2017a}.

In Fig.~\ref{fig:fdwide} we compare the magnetic field dependence of the THz spectra for the two polarization configurations for $E^\omega \parallel  a$ \& $H^\omega \parallel c$ (panel a) and  $E^\omega \parallel c$ \& $H^\omega \parallel  a$  (panel b) with the external magnetic field applied parallel to the $c$-axis. In a previous study, no splitting could be resolved for the lowest-lying electric-dipole active mode \textbf{1} in magnetic fields up to 7~T \cite{Kurumaji:2017a}. We were able to resolve a splitting of this mode in the spectrum at 7~T, only, and evaluated an average effective g-factor of about 0.1 using $\hbar \omega= \hbar \omega_0 \pm g_{\pm}\mu_B H$, where $g_{\pm}$ denotes the g-factors for the two branches (see Fig.~\ref{fig:fdwide}(c) and (d)). The g-factors for all branches are listed in Table~\ref{tab:gfactors}. In general, we find that the experimental splitting is not symmetric for modes \textbf{2} and \textbf{3}, which is reflected by our theoretically calculated g-factors listed in the same table. The calculated g-factors of modes \textbf{1} and \textbf{3}, however, are overestimated, while the one of mode \textbf{2} is nicely described by our model.

\begin{table}[t]
\squeezetable
\begin{ruledtabular}
\centering\footnotesize
   \begin{tabular}{cccccc}
\multirow{2}{*}{\textbf{mode}}       & \multicolumn{1}{c}{\multirow{2}{*}{\textbf{g-factor}}} & \multicolumn{2}{c}{\begin{tabular}[c]{@{}c@{}}$E^\omega\parallel a $ \\$H^\omega\parallel c$\end{tabular}} & \multicolumn{2}{c}{\begin{tabular}[c]{@{}c@{}}$E^\omega\parallel c $ \\$H^\omega\parallel a$\end{tabular}}  \\
                                     & \multicolumn{1}{c}{}                            & exp  & calc                                                                                                 & exp                  & calc                                                                                 \\ \midrule
\multirow{2}{*}{\textbf{\textbf{1}}} & $g_{-}$ & 0.09 & 0.36  &  -   & -  \\
                                     & $g_{+}$  & 0.11 & 0.34 &  -   & - \\ \midrule
\multirow{2}{*}{\textbf{2}}          & $g_{-}$                                          & 1.90 & 1.99                                                                                                 & 1.84                 & 1.99                                                                                 \\
                                     & $g_{+}$                                          & 1.97 & 2.01                                                                                                 & 2.03                 & 2.01                                                                                 \\ \midrule
\multirow{2}{*}{\textbf{3}}          & $g_{-}$                                          & 2.39 & 3.27                                                                                                 & 2.41                 & 3.27                                                                                 \\
                                     & $g_{+}$                                          & 2.26   & 3.03                                                                                                 & 2.24                 & 3.03                                                                                
\end{tabular}
   \caption{\label{tab:gfactors} Effective g-factors $g_\pm$ for modes \textbf{1}-\textbf{3} in comparison with the calculated values from our theoretical approach.}
\end{ruledtabular}
\end{table}

\subsection{Mid- and Near-infrared excitations}
\begin{figure}[t]
    \centering
    \includegraphics[width=0.9\linewidth,clip]{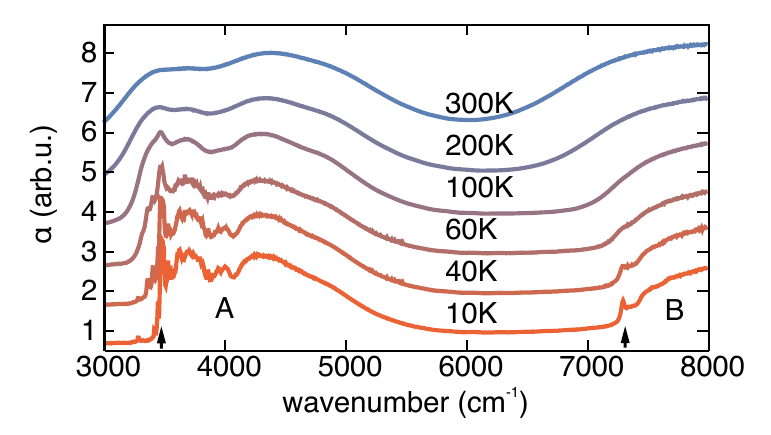}
    \caption{\label{fig:MIRtemp} Absorption spectra in \fmo{} for  polarization configuration $E^\omega \perp c$ \& $H^\omega \perp c$ at various temperatures crossing  $T_{N}=60$~K. Spectra are shifted for clarity with respect to the 10~K spectrum. Arrows mark the narrow onset excitations of the broad A- and B-bands.}
\end{figure}

Many materials with Fe$^{2+}$ ions in tetrahedrally coordinated sites exhibit electronic transitions in the mid-infrared frequency range \cite{Rudolf:2005,Ohgushi:2005,Laurita:2015,Kocsis:2018} and in \fmo{} corresponding features have been reported both in reflection \cite{Reschke:2020,Stanislavchuk:2020} and transmission experiments \cite{Park:2021}. In addition, Park \textit{et al.} ~reported the onset of a direct optical band gap at around 1~eV \cite{Park:2021}. We reinvestigated the optical properties over these frequency ranges for comparison with our theoretical approach. The temperature dependence of the absorption coefficient shown in Fig.~\ref{fig:MIRtemp} was evaluated from the transmission coefficient $T$ using $\propto 1/d \ln T$ of an $ab$-cut single crystal with a thickness of approximately 13~$\mu$m. Due to the very small sample size the reference measurement and the sample thickness possess a rather large uncertainty and, thus, we display the absorption data in arbitrary units. In agreement with previously published transmission data \cite{Park:2021}, there are two broad absorption bands called A and B in the following. Both broad bands narrow upon decreasing the temperature and develop characteristic sharp absorption features upon entering the antiferromagnetic state below 60~K. The most prominent features are the sharp peaks at the onset of band A and band B  at 3472~cm$^{-1}$ and 7298~cm$^{-1}$, respectively (see arrows in Fig.~\ref{fig:MIRtemp}). 

\begin{figure}[t]
    \centering
    \includegraphics[width=0.9\linewidth,clip]{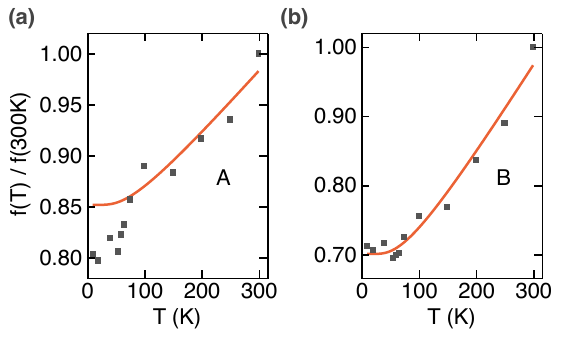}
    \caption{ \label{fig:MIRtempfit} Temperature dependence of the normalized integrated absorption strength $f(T)/f(300\,\mathrm{K})$ with a fit using $f(T)=f_c+f_0\coth{(\hbar\omega/2k_BT)}$ for (a) the A-band and (b) the B-band. }
\end{figure}

The broad absorption feature and their narrowing with decreasing temperature can be a signature of the coupling of the electronic and lattice degrees of freedom forming vibronic modes \cite{Stoneham:1975,Toyozawa:2003,Deisenhofer:2008,Laurita:2015}. A vibronic coupling has been reported in the THz regime for Zn-doped \fmo{} \cite{Csizi:2020} and for the MIR range for pure \fmo{} \cite{Park:2021} beforehand. We analyze the temperature dependence of the oscillator strength of 
modes A and B by integrating the absorbance spectra in the  range $3000-6000$~cm$^{-1}$ for band A and in the range $6000-9000$~cm$^{-1}$ for band B, respectively, and normalize to the values at 300~K. The corresponding data is shown in Fig.~\ref{fig:MIRtempfit}(a) and (b) for bands A and B, respectively. The temperature dependence of  the only partly observed broad band B can be described by $f(T)=f_c+f_0\coth{(\hbar\omega/2k_BT)}$ throughout the entire temperature regime as shown in Fig.~\ref{fig:MIRtempfit}(b). While the temperature independent term $f_c$ usually accounts for the contribution of purely electronic excitations to the oscillator strength, the second term follows the mean thermal occupation number of the phonon with energy $\hbar\omega$  which is involved in the vibronic coupling \cite{Stoneham:1975,Toyozawa:2003}. Our fit yields a phonon eigenfrequency $\omega_{\mathrm{B}} = 356\;\mathrm{cm}^{-1}$, which is larger than the value $\omega \simeq 150\;\mathrm{cm}^{-1}$ obtained  by Park \textit{et al.} \cite{Park:2021}. A simulation (not shown) using the latter value, however, also yields a reasonable parametrization of our data, indicating that the error bar of this parameter is quite large.

In the case of band A, this vibronic behavior can be used to describe the data only for temperatures above the antiferromagnetic ordering $T>75$~K with a larger phonon frequency $\omega_{\mathrm{A}} = 503$~cm$^{-1}$. Below 75~K the oscillator strength of the A-band decreases strongly and then levels off in the antiferromagnetic state at lowest temperatures. This behavior of the A band is different from the one reported by Park et al., who found a constant  oscillator strength from 4~K to room temperature \cite{Park:2021}. The origin of the A-band is attributed to the Fe$^{2+}$-ions in tetrahedral environment on the A-site \cite{Sheu:2019,Park:2021}. This is in line with previously reported optical data for materials with Fe$^{2+}$ in tetrahedral environment \cite{Rudolf:2005,Ohgushi:2005,Laurita:2015,Kocsis:2018, Strinic:2020}, and the observation that the A band absorption was found to be absent in FeZnMo$_3$O$_8$ \cite{Park:2021}, where the tetrahedral sites should have been predominantly filled by Zn ions \cite{Varret:1972,Bertrand:1975,Kurumaji:2015,Streltsov:2019,Ji:2022}.

\subsection{Nonreciprocal directional dichroism in the THz range}

\begin{figure}[t]
    \centering
    \includegraphics[width=\linewidth,clip]{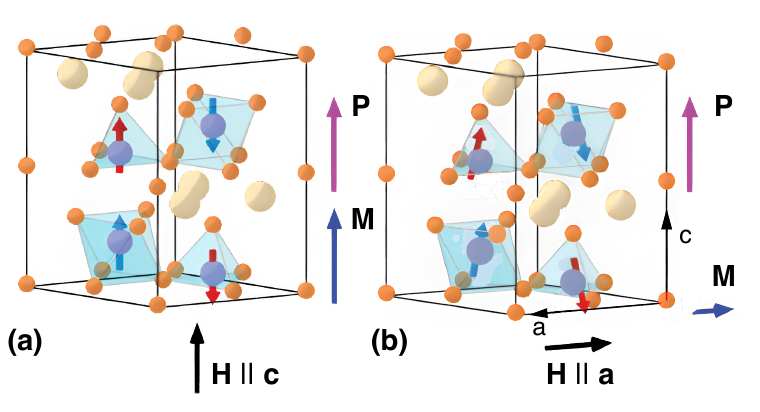}
    \caption{\label{fig:NDDStruct}   
        Comparison of the two measurement configurations (a) with the external field  applied along the polar antiferromagnetic $c$-axis, where the Zeeman-like splitting occurs as shown in Fig.~\ref{fig:fdwide}  and (b) with $\mathbf{H}\parallel a$, where non-directional dichroism can occur.}
\end{figure}

\begin{figure}[t]
    \centering
    \includegraphics[width=\linewidth,clip]{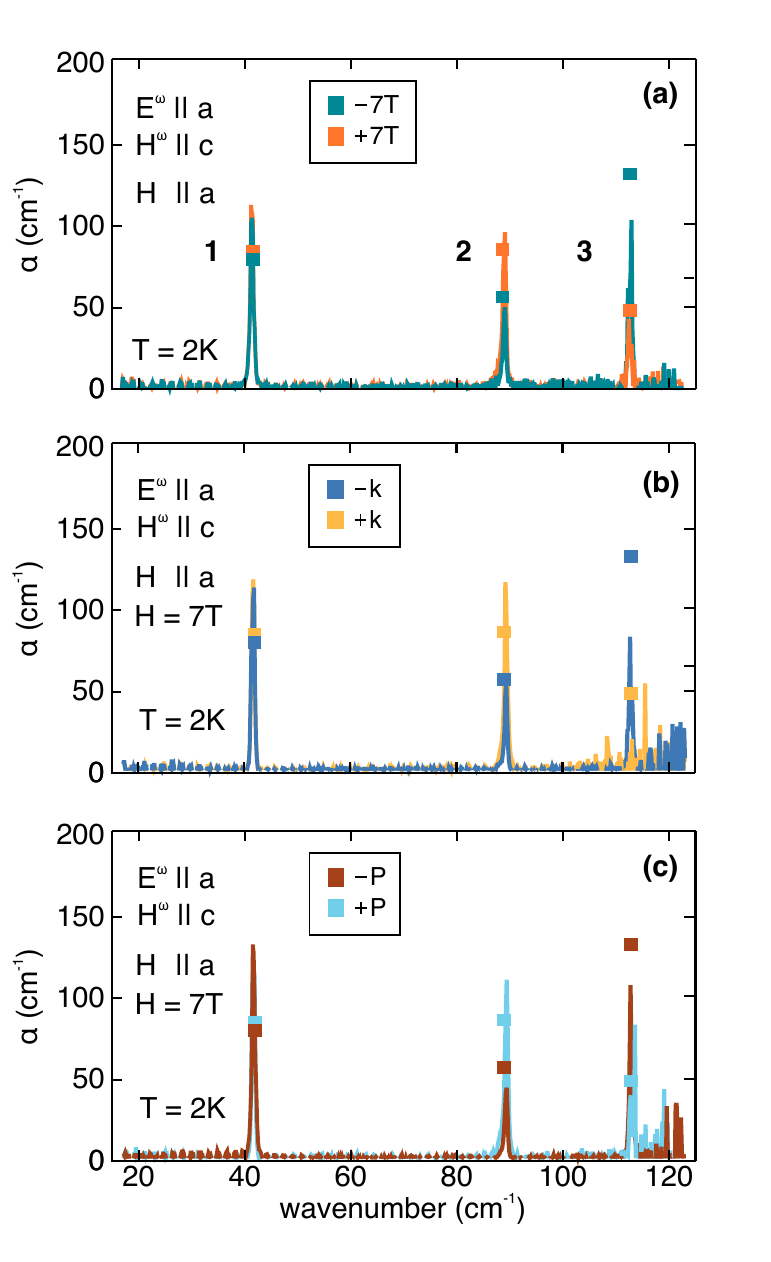}
    \caption{\label{fig:THZNccAll} Comparison of the absorption coefficient  for polarization configuration $E^\omega \parallel  a$ \& $H^\omega \parallel c$ and an external magnetic field $H=\pm 7$~T applied along the crystallographic $a$-axis. Symbols correspond to the calculated absorption coefficient at $H=\pm 7$~T as discussed in the text.}
\end{figure}

In this section we will discuss the nonreciprocal directional dichroism observed in THz absorption spectra of \fmo. In contrast to the magnetic-field effects shown in Fig.~\ref{fig:fdwide}, where the external magnetic field has been applied along the polar antiferromagnetic $c$-axis, the field will now be applied parallel to the $a$-axis and induce a magnetization component in the hexagonal $ab$-plane (see Fig.~\ref{fig:NDDStruct} for a comparison of the two configurations).
As a consequence, the $6'm'm$ hexagonal magnetic point group symmetry of the antiferromagnetic state \cite{Kurumaji:2015} will be reduced to the $2'm'm$ orthorhombic symmetry and directional dichroism for light beams propagating along and opposite to the $b$-axis is expected to occur following the trilinear product relation
$\Delta n\propto\mathbf{k}\cdot(\mathbf{P}\times\mathbf{M})$ \cite{Rikken:2002,Szaller:2013}. Here, $\mathbf{k}$ denotes the light's wave vector in propagtion direction and $\mathbf{P}$ and $\mathbf{M}$ are the polarization along the $c$-axis and the induced magnetization in the direction of the external applied field, respectively. The Maxwell equations yield the following four solutions for the refractive index $n$
\begin{eqnarray}
n^{\pm k_b}_{ac}&\approx& \sqrt{\epsilon'_{aa}\mu'_{cc}}\mp\chi'_{ca},\label{ref_index12} \\
n^{\pm k_b}_{ca}&\approx& \sqrt{\epsilon'_{cc}\mu'_{aa}}\pm\chi'_{ac}.
\label{ref_index34}
\end{eqnarray}
where $n^{\pm k_b}_{ac}$ and $n^{\pm k_b}_{ca}$ denote the refractive indices for light polarizations 
$E^\omega \parallel  a$ \& $H^\omega \parallel c$ and $E^\omega \parallel c$ \& $H^\omega \parallel  a$, respectively. The components of the electric permittivity and magnetic permeability tensor are given by $\epsilon'_{ij}$ and $\mu'_{ij}$, time-reversal odd  components of the magnetoelectric tensor are denoted as $\chi'_{ac}$ and $\chi'_{ca}$ \cite{Kezsmarki:2011}. Solutions for wave propagation in opposite directions are indicated by $\pm k_b$. The validity of the trilinear product form was explicitly confirmed for the sister compound Co$_{2}$Mo$_3$O$_8$ by reversing all three constituents - the light propagation direction reverses $\mathbf{k}$, the magnetization $\mathbf{M}$ by reversing the external magnetic field and the polarization $\mathbf{P}$ by rotating the sample by 180$^\circ$  \cite{Reschke:2020}. In Fig.~\ref{fig:THZNccAll} we show the comparison of the absorption spectra obtained by reversing all three components for \fmo{} together with the calculated absorption coefficient for the polarization configuration $E^\omega \parallel  a$ \& $H^\omega \parallel c$. Note that spectra for the reversal of propagation in Fig.~\ref{fig:THZNccAll}(b) and the ones for polarisation in Fig.~\ref{fig:THZNccAll}(c) were measured in a static field of 7T to directly compare the effects.

As expected the electric-dipole active mode \textbf{1} does not exhibit a difference in absorption upon reversal of any of the three components within experimental uncertainty in agreement with the theoretical estimate. Modes \textbf{2} and \textbf{3}, however, show clear directional dichroism effects. For mode \textbf{2} the different spectra for reversing the three components agree very well with each other and the dichroic effects are well captured by our theoretical approach. For mode \textbf{3} the results for $\pm \mathbf{k}$, which involves a complete exchange of source and detector and realignment of the optical setup, agree only qualitatively with the other two configuration, the agreement in experiment and theory is still good for $\pm 7$~T and $\pm \mathbf{P}$ and shows the success of our theoretical model.

\section{Theoretical Model}
We now discuss the theoretical model used to describe the optical excitations in \fmo{} and their behavior in external magnetic fields. The model uses a single-ion approach for the two different Fe$^{2+}$ sites with tetrahedral and octahedral coordination. In contrast to previous works using single-ion models \cite{Varret:1972, Stanislavchuk:2020}, we additionally take into account contributions of the covalent bonding and orbital overlap with the nearest ligands as well as the Coulomb interaction. We then numerically diagonalize the Hamiltonian using the $|{}^5D, M_L, M_S\rangle$ electronic multiplet basis for  both A- and B-site  Fe$^{2+}$ ions  ($3d^6,S=2,L=2$) and model the optical excitations with or without an external magnetic field. The obtained wave functions are given in the appendix in Tab.~\ref{tab:wavefunctions}.

\subsection{Effective Hamiltonian for A/B sites of Fe$^{2+}$}\label{sec:theoryMod}
To calculate the energy levels and wave functions of the Fe$^{2+}$ ions in \fmo{} we start out with the free-ion $^5D$ ground state and use the effective Hamiltonian 
 \begin{equation}
 \label{eq:effham}
     H = H_{\mathrm{CF}} + H_{\mathrm{SO}} + H_{\mathrm{SS}} + H_{\mathrm{mf}} + H_{\mathrm{Z}}.    
 \end{equation}
 The first term is the crystal field (CF) operator $H_{\mathrm{CF}} = \sum_{k,q} B_q^{(k)} C_q^{(k)}$ written in Wybourne notation with the CF parameters $B_q^{(k)}$ and the spherical tensor components $C_q^{(k)}$ \cite{Wybourne:1965}.  Because of the three-fold rotation axis present for the trigonal site symmetry of both A- and B-sites, only the CF parameters with $q=0,\pm 3$ are non-zero. In addition, calculating the matrix elements of $C_{q}^{(k)}$ within the 3\textit{d}$^{6}$ states of the Fe$^{2+}$ ground configuration requires $k$ to be even. Odd components can, however, contribute to non-zero  matrix elements in case of configuration mixing of the ground state with states of different parity (see Section \ref{sec:theoryModSubOne}). In first-order perturbation theory for both A- and B-sites only the four CF components $B_0^{(2)},B_0^{(4)},B_3^{(4)},B_{-3}^{(4)}$ will give non-zero contribution to the splitting of the ground $^{5}D$ multiplet. In the case of undistorted tetra- and octahedra it follows that $B_0^{(2)}=0$ and $B_3^{(4)}=B_{-3}^{(4)}=\sqrt{10/7}B_0^{(4)},$ and the resulting ground state at the A-site is the orbital doublet ${}^5 E$ and the excited multiplet state the orbital triplet ${}^5 T_2$, while at the B-site the order is reversed with ${}^5 T_2$ being the ground state. The actual distortions at both sites preserve $B_3^{(4)}=B_{-3}^{(4)}$, but lead to a splitting of the triplet and, usually, the three CF parameters are determined by fitting experimental data \cite{Varret:1972, Stanislavchuk:2020}. 
 
In our approach, we tried to further reduce the number of adjustable parameters by estimating $B_q^{(k)}$ using the  superposition model \cite{Eremin:1977, Newman:1987} of individual Fe-O pairs, considering the electrostatic potential of O$^{2+}$ ions as well as covalency and exchange effects:
 \begin{equation}\label{eq:superposition}
     B_{q}^{(k)} = \sum_{j} a^{(k)}(R_j)(-1)^{q} C_{-q}^{(k)} (\vartheta, \varphi)
 \end{equation}
Here, $(R_j, \vartheta_j, \varphi_j)$ denotes the position of the $j$th ligand with respect to the central ion and the $a^{(k)}(R_j)$ are intrinsic CF parameters defined for a single Fe-O$_j$ pair aligned along the local z-axis. They consist of three contributions
\begin{align}\label{eq:ak}
    a^{(k)} = a^{(k)}_{pc} + a^{(k)}_{ec} + a^{(k)}_{ex},
\end{align}
where $a^{(k)}_{pc}$ describes the electrostatic energy of $3d$ electrons in the point-charge  potential of the oxygen ligand; the second extended-charge contribution $a^{(k)}_{ec}$ takes into account the spatial distribution of the $2s$ and $2p$ electrons of the oxygen ions \cite{Garcia1984}. The third term describes covalence and exchange contributions related to the overlap of wave functions of Fe and nearest oxygen ions \cite{Malkin:1987}. Details on the calculation of the extended-charge contributions are discussed in App.~\ref{app:intCFpara}. 

The third contribution is given by
\begin{align}
\label{eq:exchangecov}
a^{(4)}_{ex}(R_j)&=\frac{9G_4e^2}{5R_j}\left(S^2_{\sigma} +S^2_{s}-\frac{4}{3}S^2_{\pi}\right),
\end{align}
introducing the parameter $G_4$ of the so-called exchange-charge model \cite{Malkin:1987} and the overlap integrals $S_{\sigma}, S_{s}, S_{\pi}$  for a separate Fe$^{2+}$-O$_j^{2-}$ pair. The calculation of these integrals is done using the Hartree-Fock wave functions of free ions \cite{Clementi:1974}, expanded in terms of Gaussian orbitals as described in \cite{Iglamov:2007} . The parameters $G_4$ describes the number of effective elementary charges located at the Fe-O bond needed to parametrize the actual chemical bond. Here, it reduces the number of adjustable parameters of the CF operator to two, namely $G_4=G$ and $B^{(2)}_{0}$. The reason for keeping $B^{(2)}_{0}$ as an adjustable parameter instead of reducing it the effective attractive charges like the other CF parameters is, that we could not achieve a convergence of our numerical calculations for its intrinsic CF parameters. While the values of the $B_q^{(k)}$ parameters lack well-established parameter ranges, the effective charge parameter $G$  corresponds to a clear microscopic picture and significantly reduces the  parameter space for comparison with experiment. For \fmo we obtained $G \simeq 11.5$ electron charges for both A- and B-site bonds. The value is a in agreement with $G\simeq 8.93$ obtained by this concept for the FeO$_4$ tetrahedral complex in FeCr$_2$O$_4$ \cite{Vasin:2021}. 

\begin{figure}[tb]
    \centering
    \includegraphics[width=0.9\linewidth,clip]{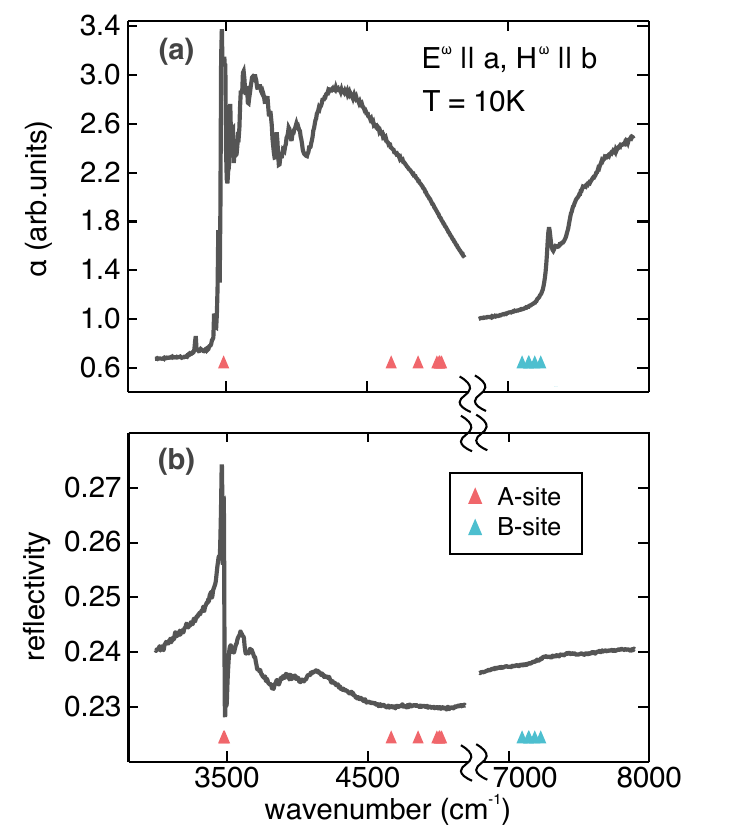}
    \caption{\label{fig:MIRTrans} Absorption and reflectivity spectrum in \fmo{} measured in the MIR/NIR frequency range with light polarization configuration $E^\omega \perp c$ \& $H^\omega \perp c$ at 10\,K and comparison with the calculated excitations as described in the text.}
\end{figure}

The $B_{q}^{(k)}$ values obtained here for \fmo{} (see Table~\ref{tab:CEFparameters} in App. \ref{app:intCFpara}) differ slightly from previous estimates reported by some of us \cite{Eremin:2022}, as we now used the observed sharp onsets of A- and B-site bands in the MIR frequency range (see Fig.~\ref{fig:MIRTrans}) to optimize the two remaining parameters $G$ and $B^{(2)}_{0}$. In comparison with other studies using single-ion approaches \cite{Varret:1972, Stanislavchuk:2020} we obtain a positive sign of the CF parameter $B_4^0$ for the A-site ions and a negative one for the B-site. We emphasize that, based on the symmetry of the spherical harmonic $Y_{0}^{(4)} \propto 9 + 20 \cos(2\vartheta) + 35 \cos(4\vartheta)$ entering $B^{(4)}_0$, this particular component of the CF operator cannot have the same sign for both tetrahedral and octahedral complexes. Hence,  the sign change is transferred also to the $B_{\pm 3}^{(4)}$ components.

The electronic states will be further split and mixed by the spin-orbit interaction $H_{\mathrm{SO}} = \sum_i \zeta_d (r_d(i)) \mathbf{l_i s_i} + \sum_i \zeta_p (r_p(i)) \mathbf{l_i s_i}$  and  the spin-spin interaction $H_{\mathrm{SS}} = \rho (\mathbf{LS})^2$. While we use the standard expression for the spin-spin interaction $H_{SS}$ with the experimentally derived  value $\rho=0.18$~cm$^{-1}$ \cite{Abragam:1970}, the spin-orbit term is non-conventional \cite{Misetich:1966} in the sense that it includes both the single-electron spin-orbit coupling $\zeta_d$ on the sites of the magnetic ions and the one on the   ligands' sites $\zeta_p$ due to charge-transfer processes. Usually, the spin-orbit coupling of $3d$ magnetic ions obtained from comparison with experimental data is reduced with respect to the corresponding free-ion values, and this reduction is often parametrized by the so-called covalency reduction factors. Here, we start with the free-ion value $\zeta_d = 404$~cm$^{-1}$~\ for Fe$^{2+}$ is  \cite{Abragam:1970} (or $\lambda \sim -101$~cm$^{-1}$ in the basis of ${}^5D$ multiplet), and $\zeta_p = 150$~cm$^{-1}$~ for the O$^{2+}$ ions \cite{Maolu:1992} and calculate the reduction of effective $\zeta_d$ directly via orbital-overlap integrals (see App.~\ref{app:SO} for details). As a result we find that the ligand contribution effectively lowers the spin-orbit coupling for both Fe sites by $\sim 10\%$, i.e.  $\lambda(\mathrm{A}) \sim -89$~cm$^{-1}$ and $\lambda(\mathrm{B}) \sim -88$~cm$^{-1}$ in the basis of the ${}^5D$ multiplet. This is in agreement with the phenomenologically assumed reduction factors for Fe$^{2+}$ ions and opens a way to further reduce the number of free parameters and compare the influence of different ligands such as oxygen, sulfur, and selenium.

We take into account the exchange (molecular) field acting on the Fe sites by the fourth term in Eq.~(\ref{eq:effham}) with the following form:
\begin{equation}\label{eq:mf}
H_{\mathrm{mf}}=I S_z  +I' S_z\ket{\psi} \bra{\psi}.
\end{equation}
The first term is the usual exchange-field operator used in previous approaches \cite{Varret:1972,Stanislavchuk:2020, Eremin:2023, Eremin:2023:P}, the second additionally takes into account possible exchange contributions from excited states and includes the projection operator $\ket{\psi} \bra{\psi}$, which selects the orbital state of the Fe$^{2+}$ ion. According to the Goodenough-Kanamori-Anderson rules, it is expected that the exchange field acting on the excited multiplet states in the energy range of 3500~cm$^{-1}$ is orbitally-dependent and differs from that of the lower-lying ${}^5 E_{g}$ states. Starting parameters $|I_\mathrm{A}|\sim 110$~cm$^{-1}$, $|I_\mathrm{B}|\sim 55$~cm$^{-1}$ (see App.~\ref{app:mf}) were taken from  early work \cite{Bertrand:1975} and successively refined to $|I_\mathrm{A}|\simeq  80$~cm$^{-1}$ and $|I_\mathrm{B}|\simeq  46$~cm$^{-1}$ by comparison with the THz excitations shown in Fig.~\ref{fig:specs}. The additional exchange field parameter $I'_\mathrm{A}\sim  - sgn(I_\mathrm{A}) \cdot 150$~cm$^{-1}$ (in combination with choosing $\Psi=\ket{M_L=0}$) was used to model the eigenfrequencies and selection rules of the onset features of the A-site MIR excitations observed in the reflectivity spectra as shown in Fig.~\ref{fig:MIRTrans} (b). The parameter $I'_\mathrm{A}$ mainly influences the high-lying A-site multiplet states (see Tab.~\ref{tab:wavefunctions}) and overcompensates the isotropic term in \eqref{eq:mf}, even changing the sign of the coupling for states with significant contributions from the $\ket{M_S,M_L=0}$ states. The B-site related high-lying excitations were modelled with $I'_B = 0$. 

The last term $H_{Z} = \mu_b (g_e \mathbf{S} + \mathbf{L}) \mathbf{H}$ in Eq.~\eqref{eq:effham} accounts for the Zeeman interaction of the orbital and spin moments with the applied magnetic field. Using the numerically obtained wave functions (see App. \ref{app:misc}), we calculated the magnetic moments at the A- and B-site and obtained $\mu_\mathrm{A} = 4.31 \mu_b$ and $\mu_\mathrm{B} = 4.95 \mu_b$, respectively. These values agree very well with the ones derived from modelling magnetization plateaus \cite{Chen:2023} and the reported averaged magnetic moment of $4.6\pm 0.2\,\mu_b$ reported by early neutron data \cite{Bertrand:1975}. As shown in Fig.~\ref{fig:fdwide}, the THz modes \textbf{2} and \textbf{3} exhibit a V-shaped splitting as a function of the external magnetic field. Our single-site Hamiltonian \eqref{eq:effham}, however, will lead to a lifting of all degeneracies in the antiferromagnetic state on all individual Fe sites. The V-shaped splitting is then a consequence of the two different Fe positions in the unit cell for both A- and B-sites (see Fig. \ref{fig:structure} (a)). The external magnetic field along the c-axis will strengthen or weaken the exchange fields on the Fe sites, depending on whether the Fe magnetic moment points in or against the direction of the external magnetic field, respectively.

\subsection{Effective light-matter coupling and calculation of optical spectra and nonreciprocal directional dichroism}\label{sec:theoryModSubOne}
In this section, we discuss the effective magnetic- and electric-dipole coupling operators, which we used to calculate the optical matrix elements and selection rules for the obtained Fe$^{2+}$-multiplet states for A- and B-site ions. The usual magnetic-dipole interaction with the magnetic field $H^{\omega}$ of the incident electromagnetic radiation can be written as
\begin{equation} \label{eq:Hmd}
    H_{\mathrm{M}}=n \mu_b (g_e \mathbf{S} + \mathbf{L})[(\mathbf{k}^{\omega}/k)\times\mathbf{E}^{\omega}]
\end{equation}
with the refractive index $n \sim 4$, which sets the relation between $E^{\omega}$ and $H^{\omega}$ in Gaussian units. 
Electric-dipole transitions are not allowed between the parity-even states of the Fe$^{2+}$ ground-state configuration.  However, the static CF components $B_{q}^{(k)}$ with odd $k$ mix the states of the 3\textit{d}$^6$ ground-state configuration with the odd-parity 3\textit{d}$^5$4\textit{p} excited-site configuration on both crystallographic sites and can lead to finite matrix elements (In principle, the same mechanism applies for centrosymmetric polyhedra in case of dynamic distortions by odd phonons \cite{Sugano:1970}). The values of the odd CF components at the octahedral B-site are comparable to the ones at the tetrahedral B-site (see Table \ref{tab:CEFparameters}) and, hence, both sites can show electric-dipole activity of equal strength. Another important source for electric-dipole activity of A- and B-site ions are virtual electron-transfer processes from the 2\textit{p} and 2\textit{s} shells of the surrounding oxygen ions (denoted as L configuration) to the 3\textit{d}-state of the iron ions, i.e. admixing 3\textit{d}$^7$L$^{-1}$ states. Both sources of electrical activity can be described by the effective one-particle operator

\begin{equation}\label{eq:rhssignle}
H_E=-\sum_{\eta, \eta^{\prime}} a_\eta^{+}\left\langle\eta\left|\left(\mathbf{d}_\mathrm{e f f} \mathbf{E}\right)\right| \eta^{\prime}\right\rangle a_{\eta^{\prime}},
\end{equation}
with an effective electric-dipole coupling $\mathbf{d}_\mathrm{e f f} \mathbf{E}$. We rewrite this expression using
\begin{equation}
a_{l m}^{+} a_{l m^{\prime}}=\sum_{k, q}(2 k+1)(-1)^{l-m}\left(\begin{array}{ccc}
l & \mathrm{k} & l \\
-\mathrm{m} & \mathrm{q} & \mathrm{m}^{\prime}
\end{array}\right) U_q^{(k)}
\end{equation}
derived in \cite{Judd:1967}. We find that in our case with $l=2$, the rank of the unit tensor operators $U_q^{(k)}$ can take on the values $0$, $2$, $4$. Since the effective electric-dipole coupling operator in Eq.~\eqref{eq:rhssignle} has the form of a scalar product,
we can express it in a more common form used for electric-dipole transitions \cite{Judd:1962, Ofelt:1962} 
\begin{align}
\label{eq:Hed}
H_{\mathrm{E}}=\sum_{k,t,p}{\{E^{\omega(1)}U^{(k)}\}}_{t}^{(p)}D_{t}^{(1k)p}.
\end{align}
Here, the brackets denote the direct product of the spherical components of the electric field $E_0^{\omega(1) }=E_z^{\omega}$, $E_{\pm 1}^{\omega(1)} = \mp (E_x^{\omega} \pm i E_y^{\omega} )/\sqrt{2}$ and the unit tensor operators $U_q^{(k)}$. The operator in Eq.~\eqref{eq:Hed} is Hermitian and is invariant under  time reversal when $1 + k + p$ is an even number. In our case with $l=2$, we only get contributions for $p=1,3,5$. The parameters $D_t^{(1 k) p}$ can be expressed as 
\begin{equation}\label{eq:finalD}
D_t^{(1 k) p}=\sum_j d^{(1 k) p}\left(R_j\right)(-1)^q C_{-t}^{(k)}\left(\vartheta_j \varphi_j\right)
\end{equation}
and allow to estimate the contributions of all Fe-O pairs in the superposition model.
We want to point that the Hamiltonian \eqref{eq:Hed} can, in principle, describe all interaction mechanisms of 3\textit{d} electrons with an electric field, including e.g. vibronic interactions. The details of the calculation  of interaction of light and the electronic states of the Fe-O pairs the obtained values of $D^{(1k)p}_{t}$ are described in App.~\ref{app:D}.

\label{sec:theoryModTwo}
To model the experimentally observed absorption spectra we use the above magnetic and electric-dipole operators and estimate the absorption coefficient using 
\begin{align}
\label{eq:intens}
    \alpha &\propto \omega_{m0}\sum_{m}|\langle m| H_{\mathrm{E}} + H_{\mathrm{M}} |0 \rangle |^2,
\end{align}
where we only considered transitions between the ground state $|0 \rangle$ and the excited state wave functions $|m \rangle$ (with energy difference $\hbar\omega_{m0}$) obtained by numerical diagonalization of the Hamiltonian given in Eq.~\eqref{eq:effham}. The corresponding theoretical values for the eigenfrequencies and intensities of the experimental excitations with or without magnetic field are shown together with the experimental data in Figs.~\ref{fig:specs}, \ref{fig:fdwide}, \ref{fig:MIRTrans}. We used a single scaling factor throughout the entire frequency range for comparison of the experimental absorption coefficients and the calculated ones using  Eq.~\eqref{eq:intens}. The scaling factor was determined by comparison with the THz data in Fig.~\ref{fig:specs}b, where the absorption maxima of all three observed modes could be reliably resolved.

The nonreciprocal directional dichroism effects due to the trilinear product $\Delta n\propto\mathbf{k}\cdot(\mathbf{P}\times\mathbf{M})$ can be calculated by using Eq.~\eqref{eq:intens} and reversing the external magnetic field in the Zeeman term, the propagation direction of light (in Eq.~\eqref{eq:Hmd}), or the polarization (corresponding to the inversion of the tetrahedral or octahedral complexes and their CF parameters). The calculated absorption values for the THz modes \textbf{1},\textbf{2} and \textbf{3} nicely agree with the experimental spectra as shown in Fig.~\ref{fig:THZNccAll}. The optical magnetoelectric susceptibility $\chi_{me}$ is determined by the interference term between the electric- and magnetic-dipole matrix elements in Eq.~\eqref{eq:intens}
\begin{align}
\label{eq:delta_intens}
    \chi_{me} &\propto \langle m| H_{\mathrm{E}}|0 \rangle \langle 0| H_{\mathrm{M}} |m\rangle + h.c.
\end{align}
Clearly, only excitations which are both electric- and magnetic-dipole active can show this cross-coupling effect. As described above the very weak dichroic effect predicted for mode \textbf{1} is not observable within experimental uncertainty, but the strong effects for modes \textbf{2} and \textbf{3} with comparable amplitudes in the electric- and magnetic-dipole channel agree nicely with the experimental data.

\section{Discussion}
The strength of our approach lies in a quantitative description of the observed non-reciprocal directional dichroism effects, the selection rules and field-dependency of the observed THz excitations and further excitations in the MIR and NIR range. The complete set of energy levels, wave functions, effective g-factors and transition probabilities from our calculations is listed in Tab.~\ref{tab:wavefunctions} and can serve as a basis for comparison with future studies of the magnetic-field dependence of excitations in the FIR and MIR regimes. 

The calculation of energy-dispersion relations of the optical modes, which are inherent to the excitonic character of the electronic and spin excitations in dielectric or polar antiferromagnets such as \fmo{} \cite{Petrov:1971,Eremenko:1992} is beyond our advanced single-ion model. While approaches tackling the calculation of dispersive Frenkel excitons in antiferromagnets are rare \cite{Kopteva:2022}, the single-ion approach is, however, still the most straightforward way to describe optical excitations in insulators or semiconductors such as \fmo{} at $q\approx 0$ \cite{Loudon:1968}, and as shown here also a successful approach.

There are, however, features of the optical excitation spectra in \fmo, which reveal the cross-coupling of lattice, electronic and spin degrees of freedom and are not described within our model:\\
The vibronic coupling of the electronic excitations related to the broad absorption band A in the MIR frequency range (see Figs.~\ref{fig:MIRtemp}, \ref{fig:MIRtempfit}, \ref{fig:MIRTrans}) as well as the  interband transitions connected to hybridization effects of band B \cite{Park:2021} are clearly beyond our single-ion approach. Concerning the lowest-lying mode \textbf{1} our model predicts that the eigenfrequency of the two A-site ions in the unit cell will show a linear increase or decrease with the external magnetic field, depending on whether their magnetic moments are pointing in or against the direction of the external field, resulting in $g\approx 0.35$. Our corresponding experimental splitting with $g\approx 0.1$ is in agreement with the observation by Wu et al.~\cite{Wu:2023}, who revealed the Raman activity of a mode coinciding in frequency with mode \textbf{1} at 41~cm$^{-1}$and a linear splitting with $g=0.11$, but it is  overestimated by our calculation (see Tab.~\ref{tab:gfactors}). In contrast to our assignment of mode \textbf{1} to be of electronic origin,
Wu et al.~\cite{Wu:2023} attribute the mode to a doubly degenerate Raman-active phonon with a calculated eigenfrequency of about 58~cm$^{-1}$, which becomes electric-dipole active as a consequence of the magnetic point-group symmetry of the antiferromagnetic phase \cite{Wu:2023}. As a consequence all predicted 13 Raman-active $E_2$ modes for the paramagnetic phase should become infrared-active in the antiferromagnetic phase and have to be compared with previously reported new infrared-active modes in the FIR range \cite{Stanislavchuk:2020,Reschke:2020}. The splitting of the putative hybridized phonon mode \textbf{1} in magnetic field is interpreted in Ref.~\onlinecite{Wu:2023} as a feature of magnon-phonon coupling to mode \textbf{2}, which is interpreted as the lowest-lying magnon or electronic excitation. The reported g-factor of the Raman mode corresponding to our mode \textbf{2} is $g_\pm= 2.0$ in agreement with our experiments and calculations. Mode \textbf{3} was also reported as a Raman-active electronic or magnon excitation with $g_\pm= 2.4$ as in a our experiments, but overestimated by our theory  (see Tab.~\ref{tab:gfactors}).

Recent neutron scattering results reported a highly dispersive mode with a $\Gamma$-point frequency coinciding with the eigenfrequency of mode \textbf{1}, which both phonon and spin-excitation-like features \cite{Wu:2023,Bao:2023}. For modes \textbf{2} and \textbf{3} the neutron scattering study revealed hybridization effects of spin and lattice excitations \cite{Bao:2023} away from the Brillouin-zone centre, hence, our description of these modes as electronic excitations of the Fe$^{2+}$ ions remains valid as the hybridization close to $q\approx 0$ is negligible. In the light of the Raman und neutron scattering results in Refs.~\onlinecite{Wu:2023,Bao:2023}, mode \textbf{2} is coupled to the lowest-lying Raman-active phonon \cite{Wu:2023} to generate the splitting of mode \textbf{1} and it is also coupled to an acoustic phonon to produce the anti-crossing features away from the Brillouin-zone center \cite{Bao:2023}. A further route to distinguish purely electronic and magnon-phonon coupling effects could be a comparison with the isostructural compound Co$_{2}$Mo$_3$O$_8$. It exhibits the same collinear antiferromagnetic order and, consequently, should have very similar phonon modes and eigenenergies. THz and neutron scattering results have been published previously \cite{Reschke:2022}, but Raman studies have not yet been reported. Finally, we want to point out that our model also allows to reproduce the reported electric-field gradient values and isomer shifts in \fmo{} \cite{Varret:1972,Eremin:2023:P} as well as the order of magnitude of the electric polarization values \cite{Kurumaji:2015,Wang:2015, Eremin:2023:P}.

In summary, we investigated the optical excitations and the nonreciprocal directional dichroism in \fmo{} from the THz to the MIR frequency range. Experimentally, we could observe a splitting of mode \textbf{1} at 41.5~cm$^{-1}$in magnetic field of, found different selection rules for mode \textbf{2}, and identified the new IR-active mode \textbf{3} at 112.6~cm$^{-1}$. In addition, we developed a semi-empirical microscopic theory, which explains many features of the observed optical spectra in terms of the single-ion states of the Fe$^{2+}$ ions, and describes previous polarization and electric-field gradient experiments, too. This approach can be generalized to any $3$d transition-metal ion complex and facilitate an assignment of optical spectra and magnetoelectric effects.

\begin{acknowledgements}
This research was partly funded by the Deutsche
Forschungsgemeinschaft (DFG, German Research
Foundation)-TRR 360-492547816. The support via
the project ANCD 20.80009.5007.19 (Moldova) is
also acknowledged.  The work of A.~R.~N., K.~V.~V. and M.~V.~E. was supported by
the Russian Science Foundation (Project No. 19-12-00244).
\end{acknowledgements}

\appendix 
\section{Calculation of the crystal-field  parameters}
\subsection{Odd and even intrinsic CF parameters} \label{app:intCFpara}
The calculation of the electrostatic contribution to the intrinsic crystal field parameters in Eq.~\eqref{eq:ak} is discussed in the following. The point charge (pc) and extended charge (ec)  contributions were calculated  using the expressions 

\begin{eqnarray}
\label{eq:akrho1}
a^{(k)}_{pc~(l~l^\prime)} &= \frac{e^2 |Z| \langle r^k \rangle_{l l^\prime}}{R^{k+1}}\\
\label{eq:akrho2}
a^{(k)}_{ec~(l~l^\prime)}(R_j)&= \sum_{m} (2k+1)(-1)^{(l-m)} \begin{pmatrix} l & k & l^\prime \\ -m & 0 & m\end{pmatrix} \times \nonumber\\ 
& \times \frac{\langle n~ l~ m|-|e|V|n\prime~ l^\prime~ m\rangle}{\Big( l || C^{(k)} || l^\prime \Big)},
\end{eqnarray}
with no free fitting parameters involved. Here,  components with odd $k$ require $nl\neq n^\prime l^\prime$ (e.g. 3\textit{d} and 4\textit{p} shells of Fe$^{2+}$ electrons), while for even $k$ the relation $nl = n^\prime l^\prime = 3d$ must hold. The involved matrix elements are calculated for the separate  Fe$^{2+}$-O$_j^{2-}$ pairs in the local coordinate system with a common $z$-axis directed along the pair. The potential of the ligand's electrons is approximated by the expansion
\begin{align}
\label{eq:potential}
V(r)=\frac{|e|}{r}\sum_i p_ie^{-\gamma_i r^2}
\end{align}
where the origin of the coordinate system is at the nucleus of the O$_j^{2-}$-ion.  Since this potential contains also the point-charge contribution, one needs to omit the corresponding term while calculating $V(r)$ to avoid double counting. The expansion coefficients $p_i, \gamma_i$ are given  in \cite{Iglamov:2007}.

In general, \eqref{eq:akrho2} is evaluated numerically on the nearest ligands using Hartree-Fock functions. However, the numerical accuracy can be improved, if the radial part of the electron's wavefunction is written as an expansion in Gaussian-type orbits
\begin{align}
    R_{nl}(r) = \sum_i S_i^{(nl)} r^l \exp(- \alpha_i^{(nl)} r^2 )
\end{align}
Then  \eqref{eq:akrho2}  for $3d$ electrons within the ground configuration ($nl = n^\prime l^\prime$) states becomes
\begin{eqnarray}
\label{eq:kl}
 a_{ec~(3d)}^{(2)}(R) &= -\sum_{i j k}\frac{e^2 p_k S_i S_j R^2}{\xi^4} \frac{5}{4} \int_0^1 \Big[ (x^2 \alpha + \gamma)^2 \times \\
& \times \Big( 7(1-x^2)+\frac{2R^2}{\xi} (x^2 \alpha+\gamma) \Big) \Big] e^{-\frac{R^2 x^2 \alpha^2}{\xi}-\frac{R^2 \alpha \gamma}{\xi}} dx, \nonumber \\
 a_{ec~(3d)}^{(4)}(R) &= -\sum_{i j k} \frac{e^2 p_k R^4}{\xi^5} \frac{9}{2} \int_0^1 (x^2 \alpha + \gamma)^4 e^{-\frac{R^2 x^2 \alpha^2}{\xi} - \frac{R^2 \alpha \gamma}{\xi}} dx.\nonumber
\end{eqnarray}
For brevity, the indices $\alpha=\alpha_i + \alpha_j$, $\xi=\gamma+\alpha$, $\gamma=\gamma_k$ are omitted in Eqs.~\eqref{eq:kl} and \eqref{eq:kl2}. The difference between \eqref{eq:kl} and the expressions from \cite{Iglamov:2007} is that we used the correct factor $(x^2 \alpha+\gamma)$ instead of  $(x^2 \alpha-\gamma)$ used in \cite{Iglamov:2007}.

Considering the matrix elements between the closest excited configuration $3d^5 4p^1$ and the ground configuration $3d^6$ ($nl = 3d$, $n^\prime l^\prime = 4p$), we derived the equation for odd CF components as follows
\begin{eqnarray}
\label{eq:kl2}
&a_{ec~(3d~4p)}^{(1)} (R) = -\sum_{ijk} e^2 p_k S_i^{(3d)} S_j^{(4p)} \frac{3}{4} e^{-\frac{\alpha \gamma}{\xi} R^2} \frac{R}{\xi^3}  \times \nonumber\\ &\int_0^1 e^{-\frac{\alpha^2 R^2}{\xi} y^2} (\gamma^2 + \alpha y^2) \Big\{ 2R^2 \frac{1}{\xi} (\alpha y^2 + \gamma)^2 + 5(1-y^2)  \Big\} dy,\nonumber\\
&a_{ec~(3d~4p)}^{(3)} (R) = -\sum_{ijk} e^2 p_k S_i^{(3d)} S_j^{(4p)} \frac{7 R^3}{2\xi^4} e^{-\frac{\alpha \gamma}{\xi} R^2} \nonumber\\
&\times\int_0^1 (y^2 \alpha + \gamma)^3 e^{-\frac{\alpha^2 R^2}{\xi} y^2} dy.
\end{eqnarray}

The odd components of the third contribution to Eq.~\eqref{eq:ak} standing for the exchange and covalency effects are also calculated between the excited configuration and the ground term ($nl=3d$, $n^\prime l^\prime = 4p$)
\begin{eqnarray}\label{eq:aexodd}
&a_{ex~(3d~4p)}^{(k)} (R) = G_{k~(3d~4p)} \frac{e^2}{R_j} \frac{(2k+1)}{(3d||C^{(k)}||4p)} \times\nonumber\\
& \sum_{\rho,m} (-1)^{2-m} \begin{pmatrix} 2 & k & 1 \\ -m & 0 & m\end{pmatrix} S_{3dm,\rho}S_{\rho,4pm}
\end{eqnarray}
where $S_{nlm,~n\prime l\prime m\prime}$ are overlap  integrals between the wavefunction of Fe and O ions.  Our calculated values are given in Table \ref{tab:overlap}. The even components corresponding to the ground states ($nl=n^\prime l^\prime = 3d$) are already given in Eq.~\eqref{eq:exchangecov}  in the main text.

\subsection{CF parameters} \label{app:CFpara}
In order to fully determine the form of the crystal field operators
\begin{align}\label{eq:bkq}
B_q^{(k)} = \sum_j (-1)^q a^{(k)}(R_j) C_q^{(k)}(\theta_j, \phi_j)
\end{align}
one needs to consider the structural factors 
\begin{align}
C_q^{(k)}(\theta_j,\phi_j)=\sqrt{\frac{4\pi}{2k+1}}Y_{k,q}(\theta_j,\phi_j),
\end{align}
where $Y_{k,q}$ are spherical harmonics, and the sum is taken over the positions of the surrounding ligands. Since the intrinsic CF parameters in Eq.~\eqref{eq:ak} consist of three contributions as described above, we can write them separately as
\begin{align}\label{eq:Bkqexxpanded}
B_q^{(k)} = B_q^{(k)} (pc) + B_q^{(k)} (ec) + B_q^{(k)} (ex)  
\end{align}
In general the sum in Eq.~ \eqref{eq:bkq} has to be taken over all lattice ions, however, the exchange contribution \eqref{eq:exchangecov}, \eqref{eq:aexodd} as well as  \eqref{eq:akrho2} decay rapidly and, hence, we can restrict the sum to be taken over the nearest ions, only. The electrostatic interaction in the point-charge approximation $B_q^{(k)}(pc)$ is  derived directly from \eqref{eq:akrho1} and given by

\begin{align}
B_q^{(k)}(pc) = - \sum_j \frac{Z_j e^2}{R_j^{k+1}} \langle r^k \rangle_{l,l^\prime} (-1)^q C_{-q}^{(k)} (\theta_j, \phi_j),
\end{align}
where $Z_j$ is the ionic charge in the lattice ($Z_\mathrm{O}=-2$, $Z_\mathrm{Fe}=2$, $Z_\mathrm{Mo}=4$) and pairs $l,l^\prime$ are taken as $3d,4p$ and $3d,3d$ for $k=1,3$, $k=2,4$, respectively. We used $\langle r^2 \rangle_{3d,3d} = 1.393~$~a.u., $\langle r^4 \rangle_{3d,3d}=4.496~$~a.u. \cite{Abragam:1970} and  $\langle r \rangle_{3d,4p} = 0.55~$ a.u., $\langle r^3 \rangle_{3d,4p} = 3.26~$~a.u. estimated by the Hartree-Fock functions of Fe$^{3+}$ \cite{Synek1967}. 

Our evaluated values of  $B_{q}^{(k)}(pc)$ are shown in the first column of  Table \ref{tab:CEFparameters}. The values listed in parenthesis with the superscript ${[ni]}$ were calculated by considering only the nearest ions for comparison. We find drastic differences for $k=1,2$, which shows that the remote ligands play an important role. According to our estimates, the largest contribution to the electrostatic energy on  long distances comes from the Mo ions.

The $G_4$ parameter was mainly determined by comparison to transmission and reflectivity spectra measured in MIR range - Fig. \ref{fig:MIRTrans}.  The obtained value $G_4 \sim 11.5$ allowed to calculate the remaining parameters of the CF operator, which are shown in Table \ref{tab:CEFparameters} for both odd and even fields and both Fe sites. Comparing it to the literature  \cite{Brick2013, Brik2019}, it turned out to be smaller than in previously published studies on oxide complexes. This can be justified since usually the contribution to $B_{q}^{(k)}(ec)$,  $B_q^{(k)}(ex)$  is approximated with a single term in  \cite{MALKIN1987, Liu2005, Brick2013, Brik2019}. However,  this approximation  is justified only for an exchange and  partially covalence effects contribution in the crystal field energy \cite{Eremin:1977} . Therefore, we calculated $a^{(k)}_{ec}$ as well as $B_q^{(k)}(ec)$ using Hartree-Fock function and a real electrostatic potential of oxygen ion, while only $a_{ex}^{(k)}$ and $B_{q}^{(k)}(ex)$ were approximated with overlap-integrals. In this sense, the  $G$ parameters listed in \cite{Brick2013} are different from the ones used by us. An explicit separation of terms $B_{q}^{(k)}(ec)$ and $B_q^{(k)}(ex)$  allowed us to conclude  $G_{k~(3d,3d)} > G_{k^{\prime}~(3d,4p)}$, where $k$ is even and $k^\prime$ is odd. In general, the contribution $B_q^{(k)}(ex)$ also includes the energy of the covalence bond between the magnetic ion and the ligand along with overlapping effects. Here, the covalency effect is suppressed for odd components ($k^\prime=1,3$)  because of a larger energy for a charge transfer process of oxygen $2p$ electrons  to an excited configuration $3d^5 4p$ of the magnetic ion \cite{Liu2005}. Thus, we assume $G_1 = G_3 \simeq 0.5~G_4$ in our approach for odd CF components.
\begin{table}[t]
\squeezetable
\begin{ruledtabular}\begin{tabular}{ccccccc}
         &  p.c.&  $\rho$&  ex.&  Total&  Total $^\dagger$& Total $^\ddagger$         \\ \midrule
 \multicolumn{7}{c}{\bf{Tetrahedron A-site}}\\
 \midrule
 $B_0^{(1)}$& 289 (-7)$^{[ni]}$& 1945& -3153& -956& -&-\\
 $B_0^{(2)}$& -& -6394& 4541& -900$^{[free]}$& 158&233\\
 $B_0^{(3)}$& 12634 (17140)$^{[ni]}$& -34711& 30661& 6505& -&-\\
 $B_3^{(3)}$& -7172 (-8246)$^{[ni]}$& 16390& -14846& -4707& -&-\\
 $B_0^{(4)}$& 1842 (1347)$^{[ni]}$& -2148& 3760& 3586& -5507&-4148\\
 $B_3^{(4)}$& 3596 (3467)$^{[ni]}$& -5202& 9844& 8238& 6693&5967\\
 \midrule
 \multicolumn{7}{c}{\bf{Octahedron B-site}}\\
 \midrule
 $B_0^{(1)}$& 4373 (-3351)$^{[ni]}$& -90& 1531& 7232& -&-\\
 $B_0^{(2)}$& -& 788& -5602& 960$^{[free]}$& -&-\\
 $B_0^{(3)}$& -2290 (-3715)$^{[ni]}$& 7419& -6661& -2290& -&-\\
 $B_3^{(3)}$& 2279 (2187)$^{[ni]}$& -4549& 4026& 2643& -&-\\
 $B_0^{(4)}$& -3965 (-4757)$^{[ni]}$& 6272& -12333& -10026& -17570 &-7618 \\
 $B_3^{(4)}$& 4752 (4948)$^{[ni]}$& -6551& 12873& 11074& 16773&-9604\\
    \end{tabular}    \end{ruledtabular}
    \caption{
    \label{tab:CEFparameters} Different contributions to the CF parameters (in cm${}^{-1}$) for tetrahedral and octahedral positions in  \fmo{} and comparison of the total values to the ones reported in Ref.~\cite{Varret:1972} $^{\dagger}$ and
    Ref.+\cite{Stanislavchuk:2020} $(^{\ddagger})$. 
   $^{[ni]}$ indicates the contribution only from the nearest ions. The paratmeters indicated by $^{[free]}$ were estimated using experimental data due to the numerical instability of the initial estimates.
    }
\end{table}

The reason for the relatively low magnitude of $B_{0}^{(4)}(A)$ compared to $B_{3}^{(4)}$ can be explained as follows: The radial parts decay rapidly, therefore the total magnitude is mostly defined by the z-component of the closest O-ions. In the crystallographic coordinate system, each FeO$_4$ complex is formed by one O-ion at the top along the local z-axis with $R=1.9446\,\AA$, and three at the bottom with $R=2.0040\,\AA$. Both contribute to $B_{0}^{(4)}$ almost equally, but with opposite signs, leading to a significant compensation, which is not the case for $B_{3}^{(4)}$.

To clarify the chosen coordinate system, the coordinates of  the nearest oxygen ions are given in Table \ref{tab:oxCoordinates}. The iron ion is placed at the origin of the coordinate system.
\begin{table}[t]
\squeezetable
\begin{ruledtabular}
     \begin{tabular}{c|cccc}
 
        \textnumero  &  $x$(O-Fe) & $y$(O-Fe) & $z$(O-Fe) & $R$(O-Fe)\\
 \midrule
        
        \multicolumn{5}{c}{A - \bf{Tetrahedron}}\\
        \midrule
        1 & 0 & 0 & 1.9446 & 1.9445\\
    
        2 & 1.7871 & 0 & -0.9078 & 2.0045\\
  
        3 & -0.8936 & 1.5477 & -0.9078 & 2.0045\\
 
        4 & -0.8936 & -1.5477 & -0.9078 & 2.0045\\

        \midrule
        
        \multicolumn{5}{c}{B - \bf{Octahedron}}\\
        \midrule
        1 & 0.7735 & 1.3398 & -1.5056 & 2.1588\\

        2 & -1.5471 & 0 & -1.5056 & 2.1588\\
    
        3 & 0.7735 & -1.3398 & -1.5056 & 2.1588\\

        4 & -0.8344 & -1.4452 & 1.2238 & 2.0694\\
 
        5 & 1.6688 & 0 & 1.2238 & 2.0694\\
     
        6 & -0.8344 & 1.4452 & 1.2238 & 2.0694\\
   
     \end{tabular}\end{ruledtabular}
     \caption{Coordinates of the oxygen ions around tetrahedral and octahedral positions with the Fe ions at the origin of the coordinate system in \fmo and bondlengths for the Fe-O pairs  (in {\AA}).}
     \label{tab:oxCoordinates}
 \end{table}
The values of crystalline field parameters ($A_2$, $A_4$, $D_q$), reported in previous works \cite{Varret:1972,Stanislavchuk:2020} (see Table \ref{tab:CEFparameters} marked with daggers), are recalculated for comparison using the relationships
\begin{align}
\label{eq:parametersRelationship}
    B^{(2)}_0&=-7A_2,&
    B^{(4)}_0&=7(3A_4+2D_q),&
    B^{(4)}_3&=2\sqrt{70}D_q.
\end{align}
They are also given in Table \ref{tab:CEFparameters}. As one sees, parameters obtained by us differ from the results of previous works. The main difference is that the sign of the crystal field parameter $B_4^0$ (A), in contrast to previous works, is positive. The negative sign of $B_4^0$ (A) does not correspond to the crystal structure of the fragment FeO$_4$. Note that the difference in signs for the parameter $B_3^{(4)}$ in \cite{Stanislavchuk:2020} and \cite{Varret:1972} is not important, because the orientation of the local coordinate system axes relative to the crystallographic one was not determined during the phenomenological fitting of this parameter. The orientation of the local axes adopted by us with respect to the positions of nearest oxygen ions is explained in \ref{tab:oxCoordinates}.

\section{Effects of overlapping of electron orbits on the spin-orbit coupling} \label{app:SO}
In the spin–orbit coupling operator, the most important terms are \cite{Misetich:1966}
\begin{align}\label{eq:hsoextended}
    H_\mathrm{SO}=\sum_i \zeta_d (r_d(i)) \mathbf{l_i s_i} + \sum_i \zeta_p (r_p(i)) \mathbf{l_i s_i}.
\end{align}
where $\mathbf{r}_d$ denotes the position of an electron on the magnetic ion site, and $\mathbf{r}_p$ is for the ligand's site.

The form of the Hamiltonian in Eq.~ \eqref{eq:hsoextended} is not suitable for practical calculations of the spin-orbit coupling for states of $3d$ electron at $r_d$ site, however, it can be converted to an effective single-particle operator 
\begin{equation}\label{eq:Feff}
    F_{eff}=\frac{1}{2} \sum a_\eta^{+} a_{\eta^{\prime}}\left\{
    \begin{array}{l}
    \left(\eta|f| \eta^{\prime}\right) - 2(\eta|f| \rho) \lambda_{\rho \eta^{\prime}} + \\
    + \lambda_{\eta \kappa}(\kappa|f| \rho) \lambda_{\rho \eta^{\prime}} + \\
    +(\eta|f| \xi)\left[S_{\xi \rho} S_{\rho \eta^{\prime}} -\gamma_{\xi \rho} \gamma_{\rho \eta^{\prime}}\right]
    \end{array}
    \right\} + c.c.,
\end{equation}
considering the superposition of the wave functions of the magnetic ion and the ligand as discussed in \cite{Eremin:2023:P}. Here, $f$ can be any two-particle operator (in our case  \eqref{eq:hsoextended}). The set of quantum numbers $\eta$, $\xi$ denotes the states of $3d$ electrons of the magnetic ion, while $\rho$, $\kappa$ are for the $2p$ electrons of the ligand; $\gamma_{\eta, \rho}$ are so-called covalency parameters, taking into account a virtual charge transfer process and $\lambda_{\eta\rho} = S_{\eta \rho} + \gamma_{\eta \rho}$. Our estimates for these parameters are given in Table \ref{tab:overlap}.

Following the approach in \cite{Eremin:2023:P} after substituting \eqref{eq:hsoextended} into \eqref{eq:Feff} we obtain
\begin{align}\label{eq:so_anistoropic}
    & H_{eff}=\sum_i (\zeta_d (r_d(i)) + \Delta\zeta) \mathbf{l_i s_i} + \\
    &+ \left[2 W_{00}^{(11)} + W_{1,-1}^{(11)}+W_{-1,1}^{(11)}\right] v_1 + \\
& +\left[-\sqrt{\frac{3}{7}} W_{00}^{(13)}+\sqrt{\frac{2}{7}}\left(W_{1,-1}^{(13)} + W_{-1,1}^{(13)}\right)\right] v_2 + \\
& +\left[\sqrt{\frac{4}{7}} W_{00}^{(13)} + \sqrt{\frac{3}{14}}\left(W_{1,-1}^{(13)} + W_{-1,1}^{(13)}\right)\right] v_3 + \\
& +\left[W_{03}^{(13)} + W_{0,-3}^{(13)} + \sqrt{3} \left(W_{12}^{(13)} + W_{-1,-2}^{(13)}\right) \right] v_4,
\end{align}
where the $W_{\pi q}^{(\kappa k)}$ are double irreducible tensor operators \cite{Judd:1967}, that act similarly to the spherical tensor operators $C^{(k)}_q$ used in the CF operator expansion, but consider both angular momentum and spin states. This is a general equation for any 3\textit{d} system coordinated by ligands with a closed 2\textit{p} shell. Here the following notations are used

\begin{widetext}
  
\begin{equation}
\label{eq:vi}
\begin{aligned}
& v_1 = \sum_j \frac{1}{\sqrt{6}} \Lambda_0^{(11) 2} \left(R_j\right) C_0^{(2)}\left( \vartheta_j, \varphi_j \right), \\
& v_2 = \sum_j \Lambda_0^{(13) 2} \left(R_j\right) C_0^{(2)} \left( \vartheta_j, \varphi_j \right), \\
& v_3 = \sum_j \Lambda_0^{(13) 4} \left(R_j\right) C_0^{(4)} \left( \vartheta_j, \varphi_j \right), \\
& v_4 = \sum_j \frac{1}{2} \Lambda_0^{(13) 4}\left(R_j\right) \operatorname{Re} C_3^{(4)}\left( \vartheta_j, \varphi_j \right).
\end{aligned}
\end{equation}
where

\begin{eqnarray}
\label{eq:LambdaValues}
    &\Lambda^{(11) 2}\left(R_j\right)=-\zeta_d \sqrt{\frac{1}{30}}\left[3\left(S_\sigma^2+S_s^2-\gamma_\sigma^2\right)+S_\pi^2-\gamma_\pi^2\right]+\zeta_p \lambda_\pi \sqrt{\frac{2}{15}}\left[\lambda_\pi-\sqrt{3} \lambda_\sigma\right], \\
&\Lambda^{(11) 2}\left(R_j\right)=-\zeta_d \sqrt{\frac{1}{30}}\left[3\left(S_\sigma^2+S_s^2-\gamma_\sigma^2\right)+S_\pi^2-\gamma_\pi^2\right]+\zeta_p \lambda_\pi \sqrt{\frac{2}{15}}\left[\lambda_\pi-\sqrt{3} \lambda_\sigma\right], \\
&\left.\Lambda^{(13) 4}\left(R_j\right)=\zeta_d \frac{1}{\sqrt{35}}\left[3\left(S_\sigma^2+S_s^2-\gamma_\sigma^2\right)-S_\pi^2+\gamma_s^2\right)\right]-\zeta_p \frac{2 \lambda_\pi}{\sqrt{35}}\left[2 \lambda_\pi-\sqrt{3} \lambda_\sigma\right] .
\end{eqnarray}

\end{widetext}

As one can see, the actual symmetry of the surrounding ions is included in the  $\nu_i$ parameters, while the ground configuration of the $3d^n$ system enters only into the reduced matrix element $(d^n, ^{2S+1} L|| W^{(k \kappa)} || d^n, ^{2S+1} L)$, therefore we introduce the notations for our case

\begin{equation}
\label{eq:wi}
\begin{aligned}
& w_1 =\left\langle d^{6} {}^5{D}{}\left\|W^{(11)}\right\| d^{6} {}^5{D}{}\right\rangle v_1 , \\
& w_2=\left\langle d^{6} {}^5{D}{}\left\|W^{(13)}\right\| d^{6} {}^5{D}{}\right\rangle v_2 , \\
& w_3=\left\langle d^{6} {}^5{D}{}\left\|W^{(13)}\right\| d^{6} {}^5{D}{}\right\rangle v_3 , \\
& w_4=\left\langle d^{6} {}^5{D}{}\left\|W^{(13)}\right\| d^{6} {}^5{D}{}\right\rangle v_4
\end{aligned}
\end{equation}
and

\begin{eqnarray}
\label{eq:DeltaZeta}
\Delta \zeta_d=&\sum_j\{\frac{\zeta_d}{5}\left[S_\sigma^2+S_s^2-\gamma_\sigma^2+\frac{4}{3}\left(S_\pi^2-\gamma_\pi^2\right)\right]\\\nonumber
&+\zeta_p \frac{\lambda_\pi}{15}\left[\lambda_\pi+2 \sqrt{3} \lambda_\sigma\right]\}
\end{eqnarray}

Note that in \cite{Eremin:2023:P} the signs of the $w_i$ are erroneous. Here, we give the corrected results calculated for A- and B-sites of the FeO$_n$ complexes in Tab.~\ref{tab:wso} using $\zeta_d = 404~\mathrm{cm}^{-1}$ \cite{Abragam:1970} and $\zeta_{p} \simeq 150~\mathrm{cm}^{-1}$ \cite{Maolu:1992}.

\begin{table}
    \centering\begin{ruledtabular}

    \begin{tabular}{c|cc} 
         $w_i,~cm^{-1}$&  A-site&  B-site  \\\midrule
         $\Delta \zeta$&  -43&  -47\\ 
         $w_1$&  20&  -18\\ 
         $w_2$&  100&  87\\ 
         $w_3$&  74&  -260\\ 
         $w_4$&  -88&  135\\ 
    \end{tabular}\end{ruledtabular}
    \caption{Calculated parameters of the spin-orbit coupling operator.}
    \label{tab:wso}
\end{table}

It should be noted that the role of anisotropic corrections to operator \eqref{eq:so_anistoropic} in the formation of the fine structure of terms Fe$^{2+}$ ($^5 D$) has not been investigated before. However, as can be seen from Table \ref{tab:wso}, the values of the parameters $w_i$ proved to be comparable with the value of $\Delta \zeta_d$, which is usually introduced when the change in the spin–orbit coupling parameter for the 3\textit{d} electron as compared to its value for a free ion is taken into account phenomenologically. It is shown in \cite{Eremin:2023:P} that not only the change in parameter $\zeta$ occurs, but also considerable anisotropic corrections appear.

\section{Initial molecular exchange field estimates} \label{app:mf}
Our approximation to the molecular exchange fields $I(A/B)$  in Eq.~\eqref{eq:mf} is based on the values for the molecular fields given in Ref. \cite{Bertrand:1975}:   
The effective exchange fields $H_{A,B}$ at the two Fe sites can be expressed as 
\begin{eqnarray}
\label{eq:berthrand}
    H_{A} &= W_{AB} M_{B} + W_{AA'} M_{A'} + W_{AB'} M_{B'} \nonumber\\
    H_{B} &= W_{BA} M_{A} + W_{BB'} M_{B'} + W_{BA'} M_{A'},
\end{eqnarray}
where $W_{\alpha,\beta}$ are the exchange constants between the Fe spins on the two sublattices, which were extracted from the magnetic transition temperatures of the substitution series Fe$_x$Zn$_{2-x}$Mo$_3$O$_8$ in Ref. \cite{Bertrand:1975}.

Comparing \eqref{eq:berthrand} to our molecular-field contribution in Eq.~\eqref{eq:mf} and substituting $M_{A,B} = n_{A,B} g_{A,B}\mu_b\sqrt{S_{A,B}(S_{A,B}+1)}$, where $n_{A,B}$ is the concentration of A-site and B-site in the series Fe$_x$Zn$_{2-x}$Mo$_3$O$_8$, we found $H_{A} = I(A) \sim -110~cm^{-1}$, $H_{B} = I(B) \sim -55~cm^{-1}$. Our corrections to those values are based on the observed energies of the low-lying THz excitations shown on Fig. \ref{fig:specs} and lead to values $|I_\mathrm{A}|\simeq  80$~cm$^{-1}$ and $|I_\mathrm{B}|\simeq  46$~cm$^{-1}$.

\section{Coupling of  3\textit{d} electrons to the electric field} \label{app:D}
We can express any electric-dipole operator as follows \cite{Judd:1962, Ofelt:1962} 
\begin{align}
\label{eq:Hed3}
H_{\mathrm{E}}=\sum_{k,t,p}{\{E^{\omega(1)}U^{(k)}\}}_{t}^{(p)}D_{t}^{(1k)p}.
\end{align}
Here, the brackets denote the direct product of the spherical components of the electric field $E_0^{\omega(1) }=E_z^{\omega}$, $E_{\pm 1}^{\omega(1)} = \mp (E_x^{\omega} \pm i E_y^{\omega} )/\sqrt{2}$ and the unit tensor operators $U_q^{(k)}$, i.e.

\begin{eqnarray}
    \{E^{(1)} U^{(k)}\}_t^{(p)} &= \sqrt{2p+1} \sum_{q, q^\prime} (-1)^{1-k+t} \times \\ & \times   \left(\begin{array}{ccc}
1 & k & p \\
q & q^{\prime} & -t
\end{array}\right)  E_q^{(1)} U_{q^\prime}^{(k)}
\end{eqnarray}

where the operator $U_q^{(k)}$ is applied on each electron in the given configuration, i.e.
\begin{align}
U_q^{(k)} = \sum_i u_q^{(k)} (\mathbf{r}_i)
\end{align}
where for each of the 6 electrons the following matrix element should be calculated
\begin{eqnarray}
&\bra{l_i, ml_i, ms_i} u_q^{(k)} \ket{l_j, ml_j, ms_j}  = (-1)^{l_i - ml_i} \delta_{ms_i, ms_j} \times \\ \nonumber &\times\left(\begin{array}{ccc}
l_i & k & l_j \\ 
-ml_i & q & ml_j
\end{array}\right)
\end{eqnarray}
Note that for hole states the matrix elements change the sign.

We consider two contributions to the effective dipole-moments of Eq.~\eqref{eq:Hed3} given by
\begin{align}\label{eq:dksum}
    D_t^{(1 k) p} = D^{(1 k) p}_{t}(\mathrm{ocfe}) + D_t^{(1 k) p}(\mathrm{cov}).
\end{align}
The first takes into account odd crystal field components (ocfe) and can be written as \cite{Judd:1962}
\begin{eqnarray}
\label{eq:dcf}
D_t^{(1 k) p}(\mathrm{ocfe})=&2|e| \frac{(2 k+1)\langle r_{l l^{\prime}} \rangle}{\left|\Delta_{l l^{\prime}}\right| \sqrt{2 t+1}}\left(\frac{\varepsilon^{\prime}+1}{3}\right)\left(l|| C^{(1)} \| l^{\prime}\right)\times \nonumber\\ 
&\left(l^{\prime}\left\|C^{(t)}\| l^{\prime}\right)\left\{\begin{array}{lll}
1 & k & t \\
l & l^{\prime} & l
\end{array}\right\} B_t^{(p)}\right..
\end{eqnarray}
In our case  non-zero contributions only arise for  $p = 1, 3$, for which we have already defined all necessary crystal field components $B_q^{(k)}$ in App. \ref{app:intCFpara}.  Moreover,  $l$ and $l^\prime$ refer to the $3d$ and $4p$ shells, respectively,  and $(l || C^{(p)} || l^\prime)$ denotes the reduced matrix elements, and $\epsilon'$ is the real part of dielectric permeability in the dc limit, which was not included in the original Judd-Offelt equation \cite{Judd:1967}.    

The averages we evaluated using Hartree-Fock functions \cite{Clementi:1974} are equal to $\langle r_{pd}\rangle \simeq 0.76$~a.u. and $\langle r^3_{pd} \rangle \simeq 4.92$~a.u. Using the parameters of the odd crystal field and $|\Delta_{pd}|\simeq 70000$~cm$^{-1}$ \cite{Simonetti:1977} and $\epsilon^\prime = 9.1$ \cite{Wang:2015}, we evaluated the crystal field contribution to the $D^{(1k)p}_t$  parameters given in Table \ref{tab:dee}.

The second term defining the covalent contribution  (cov) in Eq.~\eqref{eq:dksum} has a different meaning than that used in the analysis of electric-dipole transitions in rare-earth compounds. In the works of M.F.~Ried \cite{Liu2005} and colleagues \cite{Reid1989, XiaShangDa1993} this term refers to the contribution arising from the overlapping and covalent effects on the parameters of the odd crystal field, which mixes states of $4f^N$ and $4f^{N-1}5d$ shells. We consider a similar contribution as the renormalization of parameters in the odd crystal field, mixing states of $3d^N$ and $3d^{N-1}4p$. In our case, the covalent contribution corresponds to accounting for the delocalization of the electronic density of 3\textit{d} electrons due to overlapping and covalency effects with the nearest oxygen ions. It is not related to the odd crystal field. In the case of rare-earth compounds, this effect is considered weak and is not taken into account. For calculating the covalency contribution, we expand it for individual pairs of Fe-O as

\begin{equation}
\label{eq:D1kpt}
D_t^{(1 k) p} (\mathrm{cov})=\sum_k d_{\mathrm{cov}}^{(1 k) p}\left(R_j\right)(-1)^t C_{-t}^{(p)}\left(\vartheta_j, \varphi_j\right).
\end{equation}

We follow the approach of Refs. \cite{Eremin:2019PRB,Eremin:2023:P}, where it was derived using the secondary quantization technique, and find
\begin{widetext}

\begin{equation}
\label{eq:d1kp}
d^{(1 k) p}_{\mathrm{cov}}\left(R_j\right)=-\frac{\varepsilon^{\prime}+2}{3} \sum(-1)^{1-k} \sqrt{2 p+1}
\left(\begin{array}{ccc}
1 & k & p \\
-q & q & 0
\end{array}\right) 
d_{-q, q}^{(1 k)}\left(R_j\right),
\end{equation}
where
\begin{equation}
\label{eq:d1qq}
d_{-q, q}^{(k)}=\sum_{m, m^{\prime}}(-1)^{l-m+q}(2 k+1)\left(\begin{array}{ccc}
l & k & l \\
-m & q & m^{\prime}
\end{array}\right)\left\{l m\left|r C_q^{(1)}\right| l m^{\prime}\right\}.
\end{equation}

One should note, that it follows from Eq.~\eqref{eq:d1qq} follows that $k$ must be even, and that the matrix element (curly brackets) usually vanishes for pure $3d^6$ states. Using the method of linear combinations of atomic orbitals (LCAO) to consider an admixture of the ligands' wavefunction one can write it as follows
\begin{equation}
\left\{l m\left|r C_q^{(1)}\right| l m^{\prime}\right\}=\sum_{\kappa, \rho}\left[\lambda_{m \kappa}\left(\kappa\left|r C_q^{(1)}\right| \rho\right) \lambda_{\rho m^{\prime}}-\left(\operatorname{lm}\left|r C_q^{(1)}\right| \rho\right) \lambda_{\rho m^{\prime}}-\lambda_{m \kappa}\left(\kappa\left|r C_q^{(1)}\right| l m^{\prime}\right)\right],
\end{equation}
where we assumed that 
\begin{equation}
\lambda_{m \kappa}\left(R_j\right)=\gamma_{m \kappa}\left(R_j\right)+S_{m \kappa}\left(R_j\right).
\end{equation}

\end{widetext}
Here, $S_{m \kappa}$ denote  in a short notation the  overlap integrals described in App.~\ref{app:intCFpara}, and $\gamma_{m \kappa}$ are so-called covalency parameters per one metal-ligand pair. 

The evaluated values of $D_{t}^{(1k)p}(\mathrm{cov})$ are given in Table \ref{tab:dee}. After that they were adjusted to correct the relative intensities of the low-lying THz excitation as well as to IR's reflectivity spectra (not shown).

\begin{widetext}

\begin{table*}

\caption{\label{tab:dee}Calculated parameters of the electric-dipole operator \eqref{eq:D1kpt} for both positions of  Fe$^{2+}$ in atomic units. The fourth column (\textit{adj}) in each section show the adjusted (if applicable) values based on the relative intensities of absorption lines in the THz range (Fig. \ref{fig:specs}).}
\begin{center}\begin{ruledtabular}
\begin{tabular}{c|cccccccc}

\multirow{2}{*}{$D_t^{(1k)p}$} & \multicolumn{4}{c}{\textbf{A}} & \multicolumn{4}{c}{\textbf{B}} \\

 & $\mathrm{ocfe}$& $\mathrm{cov}$& \textbf{total}& \textbf{adj}& $\mathrm{ocfe}$& $\mathrm{cov}$& \textbf{total}& \textbf{adj}\\
\midrule
$D_0^{(10)1}$ & 0 & -0.233& -0.233& & 0 & 0.064& 0.064& \\

$D_0^{(12)1}$ & 0.049& -0.300& -0.251& & -0.371& 0.0658& -0.306& \\

$D_0^{(12)3}$ & 0.050& -3.571& -3.523& & -0.018& 0.781& 0.763& \\

$D_0^{(14)3}$ & 0.349& -0.233& 0.115& & -0.123& 0.075& -0.048& \\

$D_0^{(14)5}$ & 0& 0.567& 0.567& & 0& -0.013& -0.013& \\

$D_3^{(12)3}$ & -0.036& 1.723& 1.687& 2.834& 0.062& 0.465& 0.527& 0.843\\

$D_3^{(14)3}$ & -0.252& 0.153& 0.099& & 0.142& 0.029& 0.171& \\

$D_3^{(14)5}$ & 0& 0.126& 0.126& & 0& -0.062& -0.062& \\

\end{tabular}\end{ruledtabular}
\end{center}
\end{table*}
\end{widetext}

\begin{table}[h]
    \centering\begin{ruledtabular}

    \begin{tabular}{c|cccc} 
         &  \multicolumn{2}{c}{\textbf{A}}&  \multicolumn{2}{c}{\textbf{B}}\\ \midrule
         \textbf{R}&  1.9445&  2.0045&  2.0694&  2.1588  \\
         $S_{3d0,2p0}$&  -0.0722&  -0.0680&  -0.0634&  -0.0573\\ 
         $S_{3d0,2s0}$&  -0.0771&  -0.0697&  -0.0624&  -0.0534\\ 
         $S_{3d1,2p1}$&  0.0576&  0.0523&  0.0471&  0.0408\\ 
         $S_{4p0,2p0}$&  -0.0721&-0.0679&-0.0633&-0.0575\\ 
         $S_{4p0,2s0}$&  -0.0770&-0.0697&-0.0622&-0.0537\\ 
 $S_{4p1,2p1}$& 0.0576&0.0522&0.0470&0.0410  \\ \midrule
 $\gamma_{3d0,2p0}$& -0.2440&-0.2300&-0.2440&-0.1939\\ 
 $\gamma_{3d1,2p1}$& 0.2376&0.2160&0.2376&0.1684\\ 
 $\gamma_{3d0,2s0}$& -0.0298&-0.0270&-0.0298&-0.0207\\ 
    \end{tabular}\end{ruledtabular}
    \caption{\label{tab:overlap} Calculated overlap integrals $S_{\eta \rho}$  between Fe
    $^{2+}$ and O$^{2-}$ states for A- and B- site in \fmo.  Hartree-Fock functions are from Ref.~\cite{Clementi:1974}. For the estimations of $\gamma$ at $R=1.988\AA$ we used their values determined for Ni-O and Cr-O pairs in oxides by magnetic resonance methods \cite{Freund:1974, Freund:1983}  and assumed that $\gamma_{xx}(R) \propto S_{xx}(R)$.  The radial wave function for 4\textit{p}-electron was taken as for Fe\textsuperscript{3+}(4\textit{}p) from Ref.~\cite{Synek1967}}.
\end{table}

\clearpage

\section{Wavefunctions, eigenfrequencies, g-factors, and transition matrix elements} \label{app:misc}
In Table~\ref{tab:wavefunctions} we list the wave functions obtained by the numerical diagonalization of the Hamiltonian~\eqref{eq:effham} for the Fe$^{2+}$ states of A- and B-sites in the  $|M_L, M_S\rangle$ basis. For each state we calculated the eigenfrequency the effective $g$-factors for the linear Zeeman-splitting in magnetic field and the square $T=|\langle m| H_{\mathrm{E}} + H_{\mathrm{M}} |0 \rangle |^2$ of the transition matrix element entering into the calculation of the absorption coefficient in Eq.~\eqref{eq:intens} from the corresponding ground state for the polarization configuration  $E^\omega \parallel  a$ \& $H^\omega \parallel b$. The presented absorption coefficients $\omega~T$ were  normalized to the modes \textbf{1} and \textbf{3} in THz range (see Fig. \ref{fig:specs}) for the corresponding Fe$^{2+}$ sites.

 \begin{widetext}

\begin{table*}[b]
\caption{\label{tab:wavefunctions} Calculated eigenfrequencies (in cm$^{-1}$ with 0.1~cm$^{-1}$ accuracy) and  wave-functions (with $98\%$ accuracy) for the Fe$^{2+}$ states of A- and B-sites in the  $|M_L, M_S\rangle$ basis. The last two columns show the calculated effective $g$-factors and the absorption coefficient $\omega T$ for the transition from the ground state to the corresponding excited state for the polarization configuration  $E^\omega \parallel  a$ \& $H^\omega \parallel b$. Calculated values of $\omega~T$ were normalized taking mode \textbf{1} and mode \textbf{3} (see Fig. \ref{fig:specs}) as $1.0$ for A- and B-site respectively. }
\begin{ruledtabular}
{

\tiny
\begin{tabular}{@{}lc|lllll|lllll|lllll|lllll|lllll|ll|l@{}}
\multirow{2}{*}{}  & $M_S$  & \multicolumn{5}{c}{2} & \multicolumn{5}{c}{1} & \multicolumn{5}{c}{0}    & \multicolumn{5}{c}{-1} & \multicolumn{5}{c}{-2} & \multicolumn{2}{c}{g} & $\omega$ T \\
                   & $M_L$  & \phantom{-}2 &  \phantom{-}1        & 0        & -1          & -2    & 2         & 1        & 0        & -1          & -2 & 2         & 1        & 0        & -1          & -2& 2         & 1        & 0        & -1          & -2& 2         & 1        & 0        & -1          & -2 & $g_{+}$ & $g_{-}$ &    
                                                                                                         \\\midrule
\multirow{25}{*}{A}     & 0.0  & & & & & & & & & & &0.03& & &0.04& & & &0.05& & & &0.75& & &-0.65 & 0.0 & 0.0 & 0.00\\  & 41.8  & & & & & &-0.01& & &-0.01& & & &-0.02& & & &-0.21& & &0.22&0.54& & &0.79&  & 0.3 & 0.4 & 1.00\\  & 88.6  & & & & & &0.05& & &0.06& & & &0.06& & & &0.74& & &-0.59&0.21& & &0.24&  & 2.0 & 2.0 & 2.19\\  & 106.6  & & & & & & & & & & & &-0.04& & &0.09&0.58& & &0.8& & & &0.05& &  & 2.3 & 2.3 & 0.04\\  & 169.4  &0.05& & &0.06& & & &0.06& & & &0.79& & &-0.6&0.09& & &0.05& & & & & &  & 4.2 & 4.2 & 0.04\\  & 170.2  & & &0.01& & & &0.08& & & &0.61& & &0.79& & & &0.06& & & &-0.05& & &0.04 & 4.4 & 4.4 & 0.00\\  & 233.3  & &0.12& & &-0.04&0.62& & &0.76& & & &0.06& & & &-0.07& & &0.06&-0.01& & &-0.01&  & 6.5 & 6.5 & 0.11\\  & 256.0  & & &0.05& & & &0.8& & &-0.59& & & &-0.08& & & &-0.01& & & & & & &  & 6.2 & 6.2 & 0.00\\  & 299.0  &0.65& & &0.76& & & &0.05& & & &-0.07& & &0.06&-0.01& & & & & & & & &  & 8.6 & 8.6 & 0.00\\  & 346.0  & &0.81& & &-0.57&-0.05& & &-0.12& & & &-0.01& & & &0.01& & &-0.01& & & & &  & 8.3 & 8.3 & 0.00\\  & 3481.4  & & &0.01& & & & & & &-0.01&-0.12& & &0.02& & & &0.98& & & &0.03& & &0.11 & 1.5 & 2.3 & 0.00\\  & 3483.0  & & & & & & & &-0.01& & & & & & &-0.01&-0.1& & &0.01& & & &0.99& &  & 1.1 & 0.3 & 10$^4$\\  & 3489.5  & & & & &-0.01&-0.13& & &0.03& & & &0.98& & & &0.02& & &0.13&-0.01& & & &  & 4.3 & 4.3 & 44.82\\  & 3503.4  &-0.11& & &0.03& & & &0.99& & & &0.01& & &0.12&-0.01& & & & & & &0.01& &  & 6.3 & 6.3 & 0.00\\  & 3521.3  & & &1.00& & & & & & &0.09&-0.01& & & & & & &-0.01& & & & & & &  & 8.3 & 8.3 & 0.00\\  & 4715.0  & & & & & & & & & & &0.06& & &-0.04& & & &-0.11& & & &0.65& & &0.75 & 0.4 & 0.4 & 0.00\\  & 4864.4  & &-0.01& & & &0.12& & &-0.09& & & &-0.11& & & &0.62& & &0.76&0.06& & &-0.08&  & 1.6 & 1.6 & 46.32\\  & 4998.0  &0.59& & &-0.51& & & & & & & &0.36& & &0.48&0.11& & &-0.12& & & &0.02& &  & 6.0 & 6.8 & 3.49\\  & 5014.0  & &0.04& & &0.01&-0.5& & &0.41& & & &-0.06& & & &-0.02& & &0.01&0.62& & &-0.43&  & 2.3 & 5.2 & 10$^3$\\  & 5015.2  &-0.3& & &0.26& & & &-0.05& & & &0.05& & &0.13&0.72& & &-0.54& & & &0.07& &  & 4.0 & 3.8 & 28.20\\  & 5017.8  & & &0.01& & & &-0.06& & &-0.01&0.78& & &-0.61& & & &0.1& & & &-0.03& & &-0.03 & 5.1 & 5.1 & 0.00\\  & 5019.5  & &-0.05& & &-0.02&0.62& & &-0.51& & & &0.1& & & &-0.11& & &-0.1&0.47& & &-0.32&  & 6.0 & 3.1 & 10$^3$\\  & 5031.5  &-0.46& & &0.4& & & &-0.14& & & &0.45& & &0.55&-0.28& & &0.16& & & &-0.02& &  & 5.7 & 5.0 & 30.49\\  & 5164.8  & & &-0.08& & & &0.59& & &0.8& & & &-0.06& & & &0.01& & & & & & &  & 5.4 & 5.4 & 0.00\\  & 5313.4  & &0.57& & &0.82&0.01& & &-0.06& & & &0.01& & & & & & & & & & & &  & 7.3 & 7.3 & 0.00\\    \\\midrule

 \multirow{25}{*}{B}  & 0.0  &0.82& & &0.5& & & &0.26& & & &0.06& & &-0.08& & & & & & & & & &  & 0.00 & 0. & 0.00 \\  & 112.8  & &0.03& & & &0.73& & &0.47& & & &0.40& & & &0.18& & &-0.26& & & &-0.01&  & 3.1 & 3.2 & 1.00 \\  & 165.4  & & &-0.01& & & &-0.01& & &-0.01&-0.28& & &-0.19& & & &-0.36& & & &-0.46& & &0.74 & 9.2 & 9.3 & 0.02 \\  & 239.7  & &0.03& & &-0.02&0.49& & &0.3& & & &-0.2& & & &-0.42& & &0.67& & & &0.02&  & 6.4 & 6.3 & 0.79 \\  & 246.7  &0.28& & &0.15& & & &-0.54& & & &-0.44& & &0.65&-0.01& & &0.02& & & &0.01& &  & 2.3 & 4.7 & 0.00 \\  & 247.1  & & &0.8& & & &0.35& & &-0.47&0.09& & &0.04& & & &0.01& & & &-0.02& & &0.03 & 4.4 & 2.0 & 6.08 \\  & 289.6  & & &-0.09& & & &-0.02& & &0.07&0.76& & &0.5& & & &0.26& & & &-0.15& & &0.26 & 4.9 & 4.8 & 0.01 \\  & 425.6  &-0.02& & &-0.01& & & &0.06& & & &0.01& & &0.04&0.76& & &0.52& & & &0.38& &  & 6.3 & 6.3 & 0.00 \\  & 455.2  & &-0.57& & &0.82&-0.01& & &0.03& & & &0.04& & & &-0.01& & &0.03& & & & &  & 1.8 & 1.8 & 16.72 \\  & 494.6  & & &0.67& & & &-0.4& & &0.62&-0.03& & &0.01& & & &0.02& & & & & & &  & 2.7 & 2.7 & 19.35 \\  & 546.8  &-0.19& & &-0.09& & & &0.83& & & &-0.25& & &0.44&-0.06& & &-0.02& & & &-0.03& &  & 3.7 & 3.7 & 0.00 \\  & 607.6  & &0.02& & &-0.05&-0.27& & &-0.14& & & &0.82& & & &-0.12& & &0.27&0.31& & &0.23&  & 5.4 & 5.5 & 0.53 \\  & 630.7  & &-0.01& & &0.02&0.12& & &0.06& & & &-0.36& & & &0.09& & &-0.12&0.75& & &0.52&  & 7.5 & 7.3 & 0.11 \\  & 681.4  & & & & & & & & & &-0.03&-0.36& & &-0.2& & & &0.89& & & &-0.07& & &0.17 & 6.6 & 6.5 & 0.08 \\  & 744.7  & & & & & & & &0.01& & & &-0.01& & &-0.01&-0.39& & &-0.24& & & &0.89& &  & 8.3 & 8.3 & 0.00 \\  & 7103.2  & &0.80& & &0.56&-0.12& & &0.15& & & &-0.01& & & &-0.01& & &-0.01& & & & &  & 0.9 & 0.8 & 10$^2$ \\  & 7108.0  &-0.51& & &0.85& & & &-0.03& & & &-0.07& & &-0.04& & & & & & & & & &  & 1.0 & 1.0 & 0.00 \\  & 7148.8  & & &-0.03& & & &0.83& & &0.55&-0.07& & &0.06& & & & & & & & & & &  & 2.8 & 2.8 & 20.07 \\  & 7152.4  & &-0.13& & &-0.12&-0.51& & &0.83& & & &-0.03& & & &-0.08& & &-0.05& & & &0.01&  & 3.0 & 3.0 & 38.24 \\  & 7193.2  & & & & & & &-0.05& & &-0.07&-0.54& & &0.83& & & &-0.03& & & &-0.07& & &-0.04 & 4.9 & 5.0 & 1.17 \\  & 7193.4  &-0.03& & &0.06& & & &-0.03& & & &0.83& & &0.54&0.03& & &-0.1& & & & & &  & 4.9 & 4.7 & 0.00 \\  & 7235.3  & & & & & & & & & & & &0.09& & &0.02&-0.56& & &0.83& & & &-0.03& &  & 6.9 & 6.9 & 0.00 \\  & 7235.7  & & & & &-0.01&-0.04& & &0.07& & & &-0.03& & & &0.82& & &0.52&0.11& & &-0.2&  & 6.8 & 6.8 & 0.15 \\  & 7278.4  & & & & & &-0.01& & &0.01& & & &-0.01& & & &0.19& & &0.09&-0.56& & &0.8&  & 8.8 & 8.7 & 0.00 \\  & 7280.4  & & & & & & & & & & &-0.04& & &0.06& & & &-0.03& & & &0.86& & &0.51 & 8.7 & 8.6 & 0.01 \\

\end{tabular}
}
\end{ruledtabular}
\end{table*}

\end{widetext}

\end{document}